**Large language models selectively converge with human-shared neural semantic representations**

Chen Hong [1,2], Ximing Shao [1,2], Gangyi Feng [1,2*]

1. Department of Linguistics and Modern Languages, The Chinese University of Hong Kong, Hong Kong SAR, China
2. Brain and Mind Institute, The Chinese University of Hong Kong, Hong Kong SAR, China.

* Corresponding author
Gangyi Feng
Brain and Mind Institute
Department of Linguistics and Modern Languages
The Chinese University of Hong Kong, Hong Kong SAR, China.
Email: g.feng@cuhk.edu.hk

**Funding**

The work described in this paper was supported by grants from the National Natural Science Foundation of China (Project No. 32322090 to Gangyi Feng) and the Research Grants Council of the Hong Kong Special Administrative Region, China (Project Nos. 14614221, 14612923, 14621424, and C4001-23YF to Gangyi Feng).

**Abstract**

Interpersonal communication requires building shared semantics that enable listeners to understand speakers' meanings from their unfolding language, but the dimensional structure of this shared neural representation remains unclear. Large language models (LLMs) increasingly approximate human language capability and neural responses, raising the question of whether they capture the same semantic structure shared between human brains. Here, we combined storytelling-listening pseudo-hyperscanning magnetoencephalography (MEG) with dimension-resolved interbrain encoding modeling to compare human- and LLM-derived accounts of shared neural semantic representations. Content words from the speaker's narratives were rated by humans and five recent LLMs along ten semantic dimensions (i.e., perception, motor, space, time, socialness, animacy, emotion, attention, causality, and drive). We tested whether these dimensions explained speaker-listener neural synchronization beyond acoustic and phonological features. Both human- and LLM-derived semantic spaces explained neural synchronization, but these shared semantics are better characterized as a multidimensional neural structure rather than a single global signal. These patterns also predicted individual differences in listeners' story comprehension, linking neural alignment to behavioral understanding. However, comparable overall prediction concealed systematic differences in representational geometry. Larger LLMs aligned more closely with humans in semantic structure and neural synchronization, showing greater overlap with human semantic synchronization, but this was incomplete and dimension-dependent. The largest divergences emerged for animacy, emotion, and socialness, dimensions closely tied to agency, affect, and social experience. These findings show that LLMs capture substantial components of human shared neural semantics, but their alignment is selective. Larger or more capable models improve the approximation, whereas socially and affectively grounded dimensions are captured only partially.

## Introduction

We use language to represent concepts and convey information, from simple greetings to complex ideas (Clark, 1996). Successful mutual understanding depends on establishing a shared semantic space that links the speaker's intended meaning with the listener's interpretation (Hasson et al., 2012; Jiang et al., 2021; Pickering & Garrod, 2013; Stephens et al., 2010). Central open questions concern the nature and structure of these shared semantic representations in the brain, the dimensions that constitute this shared space, how these dimensions shape neural alignment between speakers and listeners, and which dimensions are most relevant for interpersonal understanding. Large language models (LLMs) offer a timely computational tool as they generate explicit semantic representations for comparison with human semantic structure.

This human-LLM comparison is important not only for evaluating artificial models but, more importantly, for understanding the nature of human semantic representations. Human semantics are often argued to be embodied and socially grounded, shaped by sensory-motor interaction with the physical world and by social pressures associated with communication and cooperation (Brothers, 1990; Dunbar, 1998; Gallese & Lakoff, 2005). In contrast, most LLMs derive semantic structure primarily from statistical regularities in text corpora, yielding high-dimensional representations that capture distributional similarity but may not fully reflect direct grounding in perception, action and social interaction (Du et al., 2025; Kello et al., 2025; Xu et al., 2025). If LLMs approximate human shared semantics, they should not only produce similar semantic structures in ratings or embeddings but also explain similar neural responses between interlocutors' brains. Conversely, systematic mismatches may be evident in semantic dimensions that depend strongly on embodied and interpersonal experiences.

A promising way to test these possibilities is to link semantic representations derived from humans and LLMs to an interpersonal neural signature of shared understanding. Interbrain neural coupling or neural synchronization (NS) has emerged as a neuromarker of speaker-listener alignment during communication (Hari & Kujala, 2009; Hasson et al., 2012; Jiang et al., 2012; Montague et al., 2002; Stephens et al., 2010). Neural activity between speakers and listeners can align over time, particularly in speech- and language-related regions and in regions involved in mentalization and higher-level interpretation (Hasson et al., 2004; Stephens et al., 2010). This coupling can persist after accounting for shared acoustic input (Liu et al., 2020), suggesting that NS reflects higher-level shared representations rather than sensory entrainment alone. Recent model-based work further shows that linguistic content originating in the speaker's brain can be reflected in the listener's brain, and that LLM embeddings can capture this speaker-listener coupling better than syntactic or articulatory models (Zada et al., 2024). These findings suggest that LLMs have strong potential as computational tools for understanding shared linguistic representations. What remains unknown is not only to what extent LLMs predict shared neural responses, but also, importantly, whether they reproduce the dimensional organization of human-shared semantics.

One key challenge is that semantics is inherently multidimensional (Binder et al., 2009; Huth et al., 2016). When it comes to interpersonal shared representation, its structure could be multifaceted and shaped by human experiences. Social content constitutes a substantial portion of everyday conversation (Dunbar et al., 1997), and emotional content can strongly modulate communicative intention and effectiveness (Koolagudi & Rao, 2012; Kousta et al., 2011; Scherer, 2003). Sensory-motor experience shapes conceptual representations across categories (Gallese & Lakoff, 2005; Martin, 2007; Pulvermüller & Fadiga, 2010; Vigliocco et al., 2011), and temporal and spatial information is central to describing events and narratives (Dunbar et al., 1997). These semantic dimensions show consistency across individuals (Huth et al., 2012; Jung-Beeman, 2005), but it remains unclear which dimensions contribute most strongly to shared neural representations and, critically for this study, which semantic dimensions show the strongest convergence and divergence between humans and LLMs.

Theories of semantic representation make different predictions about where human-LLM convergence or divergence should arise. Embodied cognition theory argues that conceptual knowledge is grounded in sensorimotor experience, so perceptual and action systems contribute directly to meaning (Gallese & Lakoff, 2005; Martin, 2007; Pulvermüller & Fadiga, 2010; Vigliocco et al., 2011). Evidence that action-related words engage motor systems and that sensory concepts engage perceptual systems supports the view that semantics partly depends on grounded experience rather than being solely symbolic (Barsalou, 2008; Pulvermüller, 2005). This account predicts that shared semantics involve perceptual and action-related dimensions and suggests that text-trained LLMs may show less similarity with humans on grounded dimensions (Xu et al., 2025). Complementing this view, the social-brain hypothesis proposes that the human brain evolved to manage complex social environments, prioritizing social relationships, emotional cues, and intentions (Brothers, 1990; Dunbar, 1998). This account predicts that interpersonal shared semantics should strongly involve social and affective dimensions, and that socialness, animacy, and emotion may be particularly diagnostic of where LLMs converge with or diverge from human semantic representations.

In contrast, symbolic and distributional approaches suggest an alternative route by which LLMs may approximate human semantic structure. Symbolic semantic theories, such as hierarchical network models, represent meaning through relations among abstract nodes (Collins & Quillian, 1969). Distributional approaches further show that semantic structure can be learned from word co-occurrence patterns, and modern transformer-based LLMs scale this principle to large text corpora. From predictive embeddings such as Word2Vec (Mikolov et al., 2013) to state-of-the-art LLMs (e.g., the GPT family, Achiam et al. 2023; DeepSeek, Liu et al. 2024), these models learn rich semantic regularities from text corpora. At the same time, important limitations remain, especially in simulating the sensorimotor features of human concepts (Xu et al., 2025). Multimodal training that incorporates visual information can improve alignment between models and human conceptual spaces, yielding more human-like object-concept representations (Du et al., 2025; Xu et al., 2025). These findings motivate a dimension-resolved comparison in which some semantic dimensions may be well captured by LLMs, whereas others, especially those grounded in embodied, affective, or social factors, may show more persistent gaps.

This study compares humans and LLMs in their explanations of shared semantic representations indexed by speaker-listener NS. If successful understanding requires listeners to reconstruct meaning from the speaker's unfolding language, then semantic structure should account for NS beyond acoustic and phonological features (Hasson et al., 2012; Stephens et al., 2010). Building on evidence that LLM representations increasingly reflect human neural responses to language (Antonello et al., 2023; Bonnasse-Gahot & Pallier, 2024; Gao et al., 2025), we expected increasing model scale to enhance alignment with human-derived semantic structures in both behavioral ratings and NS. However, this convergence was not expected to be uniform across the semantic space. Dimensions learned from distributional text regularities may show closer correspondence with humans, whereas those more grounded in embodied, affective, or social experience might diverge. We predicted dimension-dependent human-LLM convergence, with some dimensions aligning in semantic organization and NS contribution, while others reveal gaps due to different ways they learn to form the dimensions.

To test these hypotheses, we used pseudo-hyperscanning MEG in a storytelling-listening task with one speaker and 22 listeners. The speaker narrated eight stories, which were later played to listeners while they underwent MEG scanning. This design allowed us to measure speaker-listener neural alignment with identical auditory input, representing speech production-to-comprehension alignment during narrative transmission. We operationalized semantic representations using ten semantic dimensions, including Perception, Motor, Space, Time, Socialness, Animacy, Emotion, Attention, Causality, and Drive, to obtain behavioral ratings from both humans and five LLMs spanning different model scales. We developed a new interbrain semantic encoding analysis framework based on temporal response functions (TRFs) (Crosse et al., 2016) to predict listeners' neural responses from the speaker's activity, quantifying the unique contribution of each semantic dimension to the speaker-listener NS and the dimensional structure while controlling for acoustic and phonological features.

This design enabled a dimension-resolved comparison of human- and LLM-derived semantic representations across four analytic levels. First, we tested whether LLMs reproduced the semantic rating structure observed in human behavioral judgments. Second, we examined whether LLM-derived semantic spaces explained speaker-listener NS and had spatiotemporal neural profiles comparable to those of humans. Third, we quantified which semantic dimensions converged or diverged between humans and LLMs in their NS patterns. Fourth, we assessed whether these dimension-specific NS patterns predicted individual differences in listeners' comprehension of the speaker's stories. Together, these analyses move from behavioral semantic structure to neural alignment and then to comprehension outcomes, examining the nature of shared semantics, identifying which components support neural alignment, and indicating where current LLMs approximate or depart from human shared semantic representations.

## Materials and Methods

### Participants

A total of 48 participants participated in this study, including 6 speakers (3 females;

mean age: 23.3 years; age range: 22-25 years) and 22 listeners (12 females; mean age: 25.6 years; age range: 21-37 years) in the MEG experiment (Hong et al., 2026), along with another 20 human raters (16 females; mean age: 23.6 years; age range: 20-30 years) for the behavioral rating experiment. All participants were native Mandarin speakers with normal hearing and had no history of neurological or psychiatric disorders. Participants provided written informed consent prior to the study and received compensation for their participation. The study was approved by the Joint Chinese University of Hong Kong (CUHK)-New Territories East Cluster (NTEC) Clinical Research Ethics Committee (CREC) and the Research Ethics Committee at National Taiwan University.

**Experimental Procedure**

*MEG experiment*

The experiment involved two sequential stages. The first stage was storytelling by the speaker, followed by story listening by the listeners, simulating unidirectional speaker-listener communication (Fig. 1A). Six speakers were recorded during the original MEG storytelling session. The following speaker-listener analyses used one selected speaker, whose retelling was chosen for fluency, clarity and MEG data quality; the same recording was then presented to all 22 listeners. This protocol kept the auditory input identical for all listeners and addressed MEG constraints limiting simultaneous recording.

In the storytelling stage, the speakers' task involved two phases. First, they read each of the eight stories silently, sentence by sentence (sentence length: 6-37 words, displayed for 2.4-13.9 sec each to maintain natural pacing, resulting in total reading times of 7-9.2 minutes per story). After reading each story, participants spent a few minutes in brief preparation, then spontaneously retold each story. Their retellings were recorded using a microphone at 48 kHz, 16-bit resolution. The stories were adapted from children's stories and works by Wu Nien-Jen and Hualing Nieh Engle, with modifications to their length and complexity. For subsequent story-listening sessions and analyses, only the retelling recordings and the accompanying MEG data were used.

In the story-listening stage, each listener completed eight scanning blocks, one for each of the eight stories (1182±51 words/story; 451±25 seconds/story; 60.1 minutes in total) narrated by the selected speaker. Audio stimuli were presented binaurally via air-tube earphones at 60-70 dB SPL to ensure clarity and listening comfort. During each story, a fixation cross remained on the screen to promote sustained attention and minimize eye movements. After each story, participants answered three yes/no comprehension questions via button press to confirm engagement and understanding. After completing all stories, comprehension was evaluated using a test with an additional 56 questions (7 questions per story). Listener comprehension was quantified as the overall accuracy across all 80 questions. This design allowed us to assess both online engagement and overall comprehension while maintaining a streamlined listening task. Story order was randomized across listeners.

*Human semantic rating experiment*

Based on the transcripts of the story recordings, we extracted 766 content words (360 verbs and 406 nouns). Each word was evaluated without context along ten semantic dimensions: Perception, Motor, Space, Time, Socialness, Animacy, Emotion, Attention, Causality and Drive (Fig. 1B and see Table S1 for the working definitions). This set of dimensions was selected to encompass a broad spectrum of semantically relevant properties (Binder et al., 2016). The ten semantic dimensions can be organized into two sets. The first four constitute embodied dimensions, ordered by the degree of concreteness-to-abstractness. Perception indexes sensory experience across modalities (e.g., visual, auditory, and somatosensory). Motor captures action-related content and encompasses somatic/sensorimotor components. Time and Space are included as fundamental reference frames for describing events. The remaining six constitute human-social dimensions, arranged from relatively coarse to more fine-grained. These dimensions were intended to capture social-emotional influences.

Human ratings were obtained from 20 participants who did not take part in the MEG experiment. Semantic dimensions were rated on a 7-point Likert scale (1 = very low, 7 = very high), except that Emotion was evaluated on a 13-point scale (-6 = very negative, 0 = neutral, +6 = very positive) to separate valence. Before rating, participants were provided with instructions and working definitions for each dimension (Table S1), along with examples illustrating high and low ratings. Raters then separately evaluated how well each word related to or reflected the properties of the target dimensions according to these definitions. Ratings for each word were averaged across the 20 raters and used as predictors in subsequent feature-driven interbrain encoding analyses to quantify each dimension's contribution to speaker-listener NS. To ensure consistency across dimensions in TRF estimation, emotion ratings were rescaled to a range of 1-7 for subsequent analyses.

**MEG data acquisition and preprocessing**

*Neuroimaging recordings*

Whole-head MEG data were acquired at a 1000 Hz sampling rate using a 306-channel Elekta Neuromag TRIUX system (204 planar gradiometers and 102 magnetometers) housed in a magnetically shielded and soundproof chamber at the Imaging Center for Integrated Body, Mind, and Culture Research, National Taiwan University. Structural MRIs were obtained with a 3T Siemens Trio scanner using T1-MPRAGE sequences (TR: 2.53 s, TE: 3.37 ms, FOV: 256 mm, 1 $mm^3$ isotropic voxels) to generate participant-specific head models for MEG source analysis.

*Preprocessing*

MEG recordings were processed with MaxFilter (Elekta Oy, Finland) for external noise reduction and head-movement correction. The data were then band-pass filtered between 0.3 and 45 Hz using a one-pass zero-phase FIR filter (order 6038; Kaiser-windowed sinc; 1000 Hz sampling) to suppress slow drifts and high-frequency noise. This filter setting retained frequency components pertinent to speech-related cortical responses (Cogan & Poeppel, 2011; Lewis et al., 2015) while attenuating near-DC slow drifts that can undermine the stability of preprocessing procedures such as ICA (Winkler et al., 2015). It also aligned with the band-pass ranges typically used in TRF-based MEG speech studies (Chen et al., 2023; Karunathilake et al., 2023).

Independent component analysis (Vigário et al., 2002) was used to detect and remove artifacts arising from eye movements/blinks, muscle activity, and cardiac signals. Each independent component (IC) was assessed using its spatial topography and temporal patterns. Identified artifact-related ICs (1-5 per dataset) were removed by zeroing their contributions. The remaining neural components were then projected back to sensor space to obtain cleaned MEG data.

To align the MEG signals with the speech stream, we recorded the audio not only with the microphone described above (48 kHz sampling rate) but also via a dedicated analog channel within the MEG system (1000 Hz). Although the MEG analog channel could not preserve fine-grained speech content due to its low sampling rate, it retained the acoustic envelope and was temporally locked to the MEG recordings. We therefore aligned the high-fidelity microphone recording with the MEG analog-channel audio by matching their envelopes to achieve precise temporal alignment between MEG and speech recordings. Specifically, we extracted the spectrograms of both signals and aligned them by maximizing their correlation (Kuhlen et al., 2012). Subsequently, MEG data were epoch-aligned to story durations and downsampled to 100 Hz to optimize processing efficiency.

*Source reconstruction*

Cortical source reconstruction was conducted in MNE-Python (v1.5.1) (Gramfort et al., 2014) under Python 3.11.6. Subject-specific head models were derived from T1-weighted MRI data using FreeSurfer v7.4.0 and implemented a single-layer boundary element model (BEM) (Fuchs et al., 2002). MEG sensors were co-registered to MRI by aligning anatomical fiducials (nasion and bilateral preauricular points). The forward solution computed sensor-level magnetic fields generated by cortical dipoles distributed on source grids with 5-mm spacing (8,192 vertices per hemisphere). The inverse solution used dynamic statistical parametric mapping (dSPM) (Dale et al., 2000) with loose orientation constraints (0.2), depth weighting (0.8), and a signal-to-noise ratio of 1. Individual source estimates were morphed to the fsaverage template and parcellated using Schaefer's 1000-parcel atlas (Schaefer et al., 2018). Downstream analyses were restricted to 196 parcels within speech production and comprehension networks (Fig. S2): inferior frontal gyrus (IFG; L:10/R:13), pre/postcentral gyri (Pre/PostCG; L:15/R:14), supramarginal/inferior parietal (SMG/IPL; L:28/R:19), middle/inferior temporal gyri (MTG/ITG; L:18/R:19), superior temporal/Heschl's gyri (STG/HG; L:30/R:30).

## Semantic feature extraction

*Transcription and linguistic feature labeling*

A native Mandarin speaker with transcription experience generated preliminary transcripts of eight story recordings. Two other native speakers then independently compared each transcript with the original audio and flagged any mismatches. All discrepancies were subsequently adjudicated in team consensus meetings. For subsequent annotation, the linguistic features were pre-specified, including the onsets and durations of consonants, vowels, tones, syllables, characters, and words. Trained annotators labeled these features in Praat (Boersma, 2011) following a standardized feature-labeling protocol. Furthermore, one acoustic feature (combination of acoustic envelopes and onsets), two phonetic features (consonant onsets and vowel onsets), two pitch features (F0 height and change) and one tonal feature (tone categories) were included as features of non-interest in the following analysis of semantic features (see Hong et al. 2026 for detailed speech feature extraction procedure).

*Dimensional semantic ratings of large language models (LLMs)*

Five large language models (LLMs; GPT3.5, DeepSeek-v3, GPT4, GPT4o, and GPT4.1) were used for semantic ratings on the 766 content words (Fig. 1B). The rating prompt for LLMs contained three parts: (1) the identical instructions and working definitions of the semantic dimensions that provided to human raters; (2) the rating scale (-6 to 6 for Emotion and 1 to 7 for other dimensions); (3) the target word to be rated. Ratings were collected one word at a time via an API to minimize contextual influences on subsequent ratings. To improve the robustness of the LLM-derived ratings, the full procedure was repeated five times for each LLM, and the mean scores across the five repetitions were used in subsequent analyses. Because the exact model scale (i.e., the number of model parameters) of the GPT series hasn't been published, we ranked the models by parameter scale as GPT3.5 < DeepSeek-v3 < GPT4 < GPT4o < GPT4.1, based on third-party estimates (Achiam et al., 2023; Liu et al., 2024; Singh et al., 2023). We treated model order as an ordinal proxy for model capability rather than a direct measure of parameter scale. Analyses using this index should therefore be interpreted as testing monotonic trends across the selected model series, not as estimating a pure parameter-scaling law.

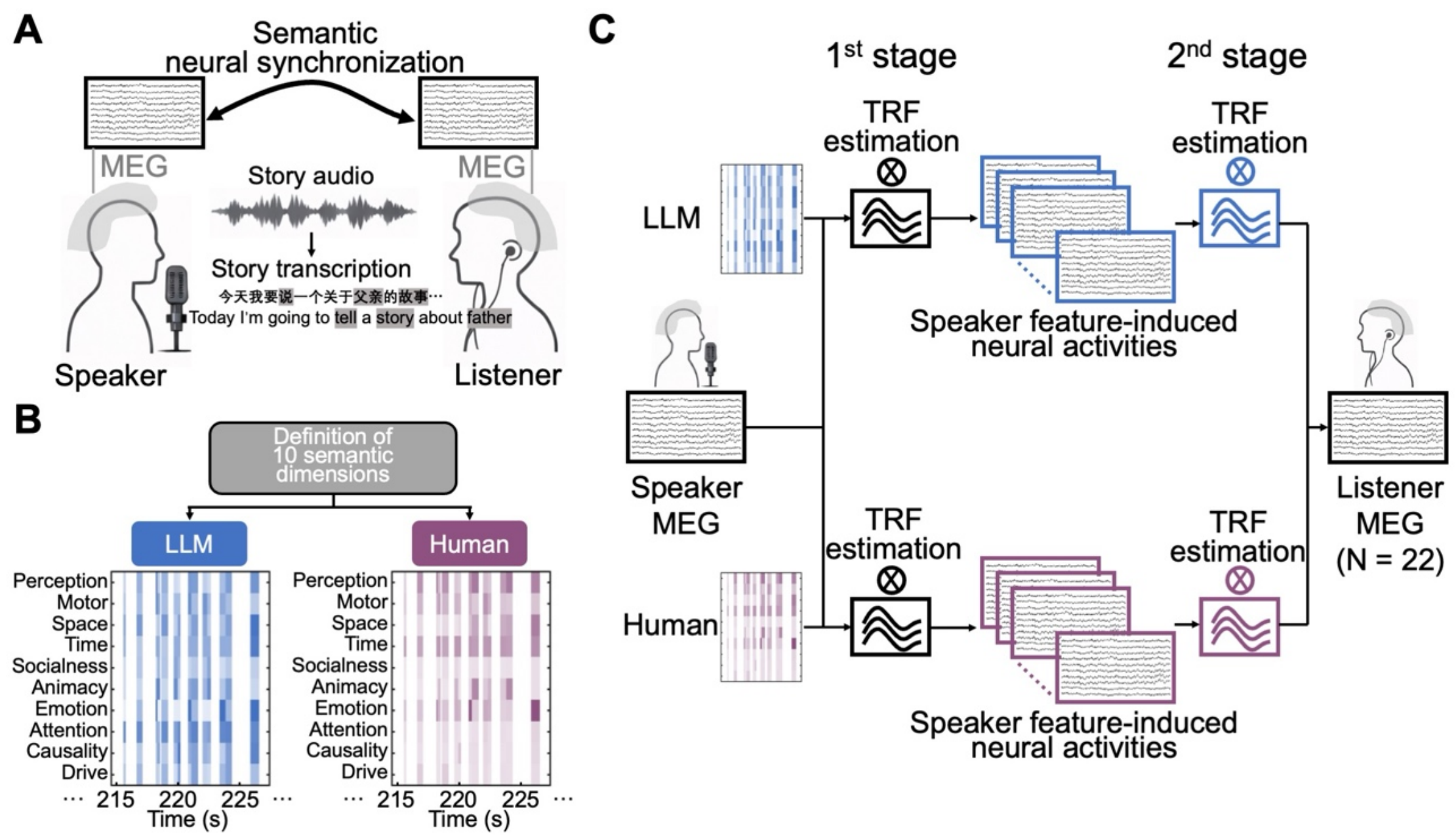

**Figure 1. Experimental design and analysis framework for speaker-listener shared neural representations across 10 semantic dimensions.** (**A**) MEG experimental setup. MEG recordings were obtained from a speaker narrating eight stories in Mandarin Chinese (example sentence shown: “今天我要说一个关于父亲的故事” / “Today I’m going to tell a story about father”). The recorded audio clips were then presented to 22 listeners, and their brain activities were recorded using MEG. All verbs and nouns were analyzed in the subsequent analyses (the shaded words in the example sentence indicate those included: “说, 父亲, 故事” / “tell, father, story”). (**B**) Semantic ratings derived from both LLMs and human raters. Semantic features were constructed by assigning, within each content word’s duration, the corresponding 10-dimensional semantic rating scores (colored areas) and setting all other time points to zero (white areas). (**C**) Two-stage feature-driven interbrain encoding modeling framework to estimate speaker-listener neural synchronization. TRF estimation was used to quantify each semantic feature in the speaker’s MEG data and to predict listeners’ neural responses. In the first stage, the semantic dimensions obtained from the LLMs and human raters were used as predictors to estimate feature-induced neural activities. In the second stage, these feature-induced neural activities served as inputs to a TRF model that predicted the listeners’ neural responses. In these TRF models, acoustic and phonetic features were included as control variables of no interest.

**Speaker-listener neural synchronization (NS) modeling**

To quantify speaker-listener NS during naturalistic speech processing, we used a multivariate temporal response function (mTRF) framework implemented in the mTRF Toolbox (Crosse et al., 2016). This framework models the linear relationship between time-varying speech-related predictors and neural responses across multiple temporal lags. Here, we adapted this approach to estimate feature-driven speaker activity in the first stage and to assess the extent to which these feature-induced speaker signals predicted listeners' neural responses in the second stage.

Semantic feature time series were constructed from content words in the story transcripts and temporally aligned with the MEG recordings. For each semantic dimension, all time points within a content word's duration were assigned that word's rating on the relevant dimension (Fig. 1B). Emotion ratings were rescaled to a 1-7 range, whereas all other dimensions retained their original 1-7 scale. Time points not corresponding to a rated content word were set to zero.

To examine how specific semantic dimensions contribute to speaker-listener NS, we used a two-stage TRF modeling approach (Fig. 1C). In the first stage, we modeled the speaker's neural encoding of individual semantic dimensions by training TRF models that used each dimension as input and the speaker's MEG responses as output. The feature set included 10 semantic features derived from LLMs and human ratings, along with control acoustic and phonology features. The TRF model is defined as:

$$y(t,n) = \sum_{\tau} \omega(\tau,n)x(t-\tau) + \varepsilon(t,n)$$

where $y(t,n)$ denotes the speaker's neural responses at the source parcel $n$ and time $t$, $x(t-\tau)$ denotes the feature at time $t-\tau$, $\omega(\tau,n)$ denotes the TRF weight at timelag $\tau$ for source parcel $n$, and $\varepsilon(t,n)$ denotes the residual error.

The TRF weights $\omega$ were estimated using ridge regression, an approach that optimized the fit between predicted and observed speaker neural activity by minimizing mean squared error while incorporating a regularization term to mitigate overfitting. To measure potential delayed and anticipatory neural responses related to speech processing, we analyzed a wide time-lag window spanning -2000 to 2000 ms. The strength of regularization (lambda) was fixed at $\lambda = 10$, providing sufficient constraint to ensure smooth TRF estimates and to minimize overfitting.

The neural signals predicted by the TRF model $\sum_{\tau} \omega(\tau,n)x(t-\tau)$ were interpreted as feature-induced neural activities, reflecting the speaker's brain responses specifically attributable to the encoding of each feature. To control for acoustic and speech features, all six features (acoustic features, consonant onset, vowel onset, pitch height, pitch change, and tone categories) were jointly incorporated into a single TRF model to produce a unified neural estimate that accounts for their variances.

In the second stage, the feature-induced speaker neural estimates served as predictors of listeners' neural responses. Specifically, for each listener parcel, we fitted TRF models in which $x$ denotes the feature-induced speaker's neural activities and $y$ denoted the listeners' neural signals, following a procedure analogous to that previously used in the first stage. The time lags were restricted to -2000 ms to 2000 ms. To estimate the unique contribution of a given semantic dimension on NS, we adopted the unique variance explained (UVE) metric, which provides a conservative quantification (Tezcan et al., 2023). We first fit a full TRF model that included both the control features and the semantic feature(s) of interest. A corresponding reduced TRF model was then conducted, which was identical except for the omission of the feature(s) of interest. UVE of the feature(s) of interest was defined as the extra variance explained by their inclusion, calculated as the variance explained by the full model minus that explained by the matched reduced model:

$$UVE = R^2_{full\ model} - R^2_{reduced\ model}$$

$$R^2(y,\hat{y}) = 1 - \frac{\sum_t (y_t - \hat{y}_t)^2}{\sum_t (y_t - \bar{y})^2}$$

where $\hat{y}_t$ denoted the prediction of the signal at the time point $t$ and $y_t$ denoted the corresponding true value; $\bar{y}$ represents the mean of $y_t$ across all time points. This procedure attributes to each feature only the variance uniquely linked to it, thereby quantifying its specific contribution to speaker-listener NS.

To assess overall spatiotemporal NS patterns associated with LLMs and human ratings, the full model included control features and all 10 dimensions of each LLM or human rating, whereas the reduced model included only the control features. To analyze dimensional NS patterns, the full model included the control features and one of the 10 dimensions from each LLM or human rating, with the reduced model unchanged. To characterize human-LLM shared and distinct dimensional NS patterns, the full model combined all 10 human rating dimensions (aggregated into a single neural estimate in the first stage) with one LLM dimension. The reduced model excluded the LLM dimension. To quantify the degree of convergence between each LLM dimension and human ratings, we defined overlap as the ratio of the variance explained ($R^2$) by the human-only reduced model to that explained by the full human-plus-LLM model. This measure indicates how much of the model's explanatory power, which includes LLMs, was already captured by human-derived semantic dimensions. Values closer to 1 suggest greater overlap, while lower values indicate additional LLM-specific explanatory variance.

The source-space analysis was restricted to speech- and language-related parcels that had previously shown robust involvement in speech production, comprehension, and interbrain NS (Hong et al., 2026). We selected 157 parcels for the speaker and 190 parcels for the listeners, as these speaker/listener parcels exhibited significant NS with at least one listener/speaker parcel in the direct speaker-listener NS analysis (Hong et al., 2026). The pipeline proceeded in two steps. First, for each speaker parcel, we estimated feature-induced neural activities separately. Second, we fitted a TRF that used the feature-induced neural activities from a given speaker parcel as input and predicted the neural responses in each of the listeners' 190 parcels. Repeating this modeling for all speaker parcels produced a UVE matrix of size 157 × 190, along with TRF weights of size 10 features × 157 × 190 × 401 time points (-2000 to 2000 ms at 100 Hz). Both stages underwent a leave-one-story-out cross-validation procedure to prevent overfitting.

To emphasize region-level NS structure, parcel-level results were aggregated into ten bilateral anatomical regions covering the inferior frontal gyrus (IFG), precentral/postcentral gyri (Pre/PostCG), supramarginal/inferior parietal cortex (IPL), middle/inferior temporal gyri (M/ITG), and superior temporal/Heschl's gyri (STG/HG). This regional pooling produced arrays of size 10 × 10 for UVE and 10 dimensions × 10 × 10 × 401 for the TRF weights.

**Classification of NS patterns of LLMs and human ratings**

To test whether human- and LLM-derived semantic NS patterns differed in their multivariate organization beyond summary measures such as overall UVE or peak latency, we performed cross-validated classification analyses. Each model condition, including the five LLMs and human ratings, was represented by listener-specific spatial and temporal NS patterns.

For spatial classification, the input features were the vectorized UVE values across the 100 speaker-listener region pairs. For temporal classification, the input features were the TRF-weight time series across 401 lags from -2000 to 2000 ms, averaged across the ten semantic dimensions and the 100 speaker-listener region pairs. Thus, the spatial analysis tested whether models differed in the distribution of semantic NS across region pairs, whereas the temporal analysis tested whether models differed in their time-lag profiles.

For each pair of model conditions, we trained a linear discriminant analysis (LDA) classifier to distinguish their NS patterns. Classification performance was evaluated using leave-one-listener-out cross-validation. In each fold, all data from one listener were held out for testing, and the classifier was trained on the remaining listeners. This procedure ensured that classification accuracy reflected generalization to unseen listeners. To assess statistical significance, we generated a null distribution by repeating the full classification procedure 10,000 times with permuted model-condition labels. Observed classification accuracy was then compared against this permutation-based null distribution.

**Representation similarity analysis**

Representation similarity analysis (RSA) (Kriegeskorte et al., 2008) was used to quantify the correspondence between the dimensional structure of semantic ratings and semantic NS patterns across LLMs and human ratings. For each model condition, including the five LLMs and human ratings, we organized the data into dimension-wise matrices. For the rating analysis, the matrix contained semantic ratings for 766 content words across 10 dimensions, yielding a 766×10 word-by-dimension matrix. For the NS analysis, the matrix contained dimension-specific NS values across 100 speaker-listener region pairs, yielding a 100×10 region-pair-by-dimension matrix.

We then constructed representational dissimilarity matrices (RDMs) by computing pairwise dissimilarities between the ten semantic dimensions within each model. For rating patterns, dissimilarity was defined as 1- cosine similarity between the word-wise rating vectors of two dimensions. For NS spatial patterns, dissimilarity was defined as 1 - Spearman's correlation between the region-pair-wise NS vectors of two dimensions. This procedure produced a 10×10 dimensional RDM for each model. To compare representational structure between models, we vectorized the upper triangular elements of each RDM, excluding the diagonal, and computed Spearman's correlations between the vectorized RDMs. This yielded a model-by-model similarity matrix indexing the correspondence of dimensional representational geometry across humans and LLMs.

To examine shared neural representations from human and LLM-derived semantics, we performed searchlight RSA on dimension-resolved TRF-weight patterns. For each model, data included TRF weights for 10 semantic dimensions across 100 region pairs and 401 lags (-2000 to 2000 ms). Temporal RSA assessed representational geometry at each lag, averaging TRF weights within a 200-ms window around each lag. This produced a vector of TRF weights for each semantic dimension across region pairs, from which dissimilarities were calculated with 1 - Spearman's correlation, forming a 10×10 RDM per lag and model. Spatial RSA evaluated each region pair separately; TRF weights across all 401 lags served as features, and dissimilarities were again computed with 1 - Spearman's correlation, resulting in a 10×10 RDM per region pair and condition.

Human-LLM shared semantic representations were quantified by comparing each LLM's RDM with the corresponding human-rating RDM. Specifically, the upper-triangular elements of the human and LLM RDMs were vectorized and correlated using Spearman's correlation. This procedure produced time-resolved human-LLM similarity profiles for the temporal searchlight analysis and region-pairwise human-LLM similarity maps for the spatial searchlight analysis.

**Multidimensional scaling and Procrustes alignment**

To characterize the relationship between the dimensional NS patterns derived from LLMs and from human ratings (10 dimensions × 100 region pairs), we first embedded the NS patterns into a two-dimensional space (10 dimensions × 2 axes) via multidimensional scaling (MDS). We then applied Procrustes alignment (Kendall, 1989) to remove non-shape dissimilarities due to translation, rotation, and isotropic scaling, thereby highlighting the shape correspondence between their dimensional patterns. The objective function minimized in the Procrustes procedure can be written as follows:

$$\min_{\rho,\mathcal{R},\gamma} \|X - (\rho Y \mathcal{R} + 1\gamma^T)\|^2$$

where $X$ is the template NS patterns (10 × 2) calculated by averaging across all speaker-listener pairs and models (5 LLMs and human ratings), $Y$ represents the reduced NS patterns (10 × 2) of each speaker-listener pair and each model, $\rho$ is a scalar isotropic scaling factor, $\mathcal{R}$ denotes an orthogonal matrix (2 × 2) representing rotation or reflection, $\gamma$ is a translation vector (2 × 1), and 1 refers to a column vector (10 × 1) of all ones. This indicates that each NS pattern was aligned to a template derived from averaging across all models and speaker-listener pairs. The distance between two NS patterns is defined as the sum of the squared differences between them.

**Statistical analysis**

To assess the statistical significance of each LLM's or human's ability to explain NS, we compared its performance against a null (control) model based on randomized ratings. Specifically, for each word, we replaced the original rating with a random value uniformly sampled from its original range and then quantified the extent to which these randomized ratings explained NS using the same analysis pipeline as for the empirical ratings. This null-model construction preserves the temporal structure of the word sequence while selectively disrupting the semantic information carried by the ratings, thereby providing a more direct test of whether semantic content per se accounts for NS.

To identify the significant time windows of TRF weights (10 dimensions × 100 region pairs × 401 timelags for each model and each subject), they were averaged across region pairs and compared with baseline TRF weights derived from TRF models trained by surrogate speaker signals. These surrogates were created by phase-randomizing the original speaker signal while retaining its amplitude spectrum and inter-channel correlation structure (Prichard & Theiler, 1994). This approach provides a stricter, physiologically motivated estimate of the time-lag interval exhibiting significant effects (Váša & Mišić, 2022).

To assess the effect of model scale, we fitted a linear regression model for each neural or behavioral representation:

$$Y = \beta_0 + \beta_1 x + \varepsilon$$

where $Y$ denotes the neural or behavioral representational measure; $x$ denotes the model-scale index ranging from 1 to 5 (corresponding to GPT3.5, DeepSeek-v3, GPT4, GPT4o, and GPT4.1), and captures the linear effect of model scale; and $\varepsilon$ denotes the residual error. The overall statistical significance of the fitted linear model was assessed using an F-test comparing it to a constant null model (intercept-only).

To assess model effect, we conducted a one-way ANOVA with model (6 levels: GPT3.5, DeepSeek-v3, GPT4, GPT4o, GPT4.1, Human) as a within-subject factor. To assess the LLM effect, we conducted a one-way ANOVA with LLM (5 levels: GPT3.5,

DeepSeek-v3, GPT4, GPT4o, and GPT4.1) as a within-subject factor. To assess the effect of semantic dimension, we conducted a one-way ANOVA with dimension (10 levels: Perception, Motor, Space, Time, Socialness, Animacy, Emotion, Attention, Causality and Drive) as a within-subjects factor. To assess LLM and dimension effect, we conducted a two-way ANOVA with LLM (5 levels: GPT3.5, DeepSeek-v3, GPT4, GPT4o, and GPT4.1) and dimension (10 levels: Perception, Motor, Space, Time, Socialness, Animacy, Emotion, Attention, Causality, and Drive) as within-subject factors. To correct for multiple comparisons, we applied the false discovery rate (FDR) correction (Benjamini & Hochberg, 1995).

**Predictive modeling of listeners' story comprehension**

To test whether NS patterns derived from LLMs and human ratings were related to listeners' comprehension outcomes, we built and evaluated models that predict comprehension scores, defined as accuracy on comprehension questions administered after story listening. One participant (72.5% accuracy) was excluded from this analysis because 72.5% was more than three standard deviations below the group mean (mean = 92.2%, SD = 5.8%). Predictors were the dimensional NS patterns spanning all speaker-listener region pairs (10 dimensions × 100 region pairs). For each LLM and for the human-rating condition, we fit a 10-fold cross-validated linear model estimated by ordinary least squares with ridge regularization. Predictive performance was summarized using the coefficient of determination ($R^2$), which indicates the fraction of variance in comprehension accuracy explained by the NS predictors. To assess robustness, we repeated the randomized 10-fold regression procedure 10,000 times and used the median $R^2$ across iterations as the performance estimate. This value was evaluated against a null distribution obtained by permuting the participant-wise mapping between NS patterns and behavioral accuracy. Dimension-specific prediction was computed with the same pipeline, except that the predictor set was restricted to the single-dimension NS vector for each dimension (1 × 100 region pairs).

## Results

**LLMs approximated human behavioral semantic structure with increasing model scale**

We first confirmed that the stories were reliably understood and provided a stable basis for examining shared semantic representations. Listeners showed high comprehension accuracy for the eight stories (mean = 92.2%, s.d. = 5.8%, range = 72.5-98.8%), significantly above chance ($t_{(21)} = 34.0$, $p = 7.6 \times 10^{-20}$).

We then examined the extent to which LLMs and humans assigned similar semantic structure to the content words in the 10-dimensional semantic space. Semantic ratings were obtained for 766 content words from 20 human raters and from five LLMs (GPT3.5, DeepSeek-v3, GPT4, GPT4o, and GPT4.1) (see distributions of the ratings in Fig. S1). Repeating the LLM rating procedure with identical prompts five times showed expected high internal consistency (pairwise Pearson's $r = 0.80$ for GPT3.5, 0.75 for DeepSeek-v3, 0.85 for GPT4, 0.89 for GPT4o, and 0.87 for GPT4.1).

At the semantic structure level, representational dissimilarity analysis (RSA; see Fig. 2A for the procedure) revealed broad similarity between LLMs and humans, as well as dimension-specific differences between them (Fig. 2A&B). In the LLM RDMs, Animacy was relatively more separated from other dimensions, whereas in human RDMs, Space was the most prominent distinction (Fig. 2B). RSA between LLM and human RDMs varied significantly across models ($F_{(4,95)} = 18.3$, $p = 3.6 \times 10^{-11}$), with GPT4, GPT4o, GPT4.1, and DeepSeek-v3 all exhibiting significantly higher similarity to human than GPT3.5 (post-hoc $p$ values from $9.9 \times 10^{-9}$ to $3.3 \times 10^{-7}$) (Fig. 2C). Larger models were associated with higher similarity to human semantic structure ($R^2 = 0.158$, $F_{(1,98)} = 19.6$, $p = 2.5 \times 10^{-5}$) (Fig. 2C).

At the individual-dimension level, most LLM-human correlations based on each dimension's ratings were significant, except for Causality in GPT3.5 (permutation $p$ = 0.054) (Fig. 2D). The LLM-human correlations were significantly correlated to LLM scales, with $F$-values ranging from 9.8 to 62.7 and $p$-values between $3.8 \times 10^{-12}$ and $2.3 \times 10^{-3}$. The only exceptions are Emotion ($F_{(1,98)} = 3.1$, $p = 0.083$) and Animacy ($F_{(1,98)} = 1.6$, $p = 0.211$), as shown in Fig. 2D. Larger LLMs more closely approximated the structure of human semantic behavior, but not uniformly across dimensions.

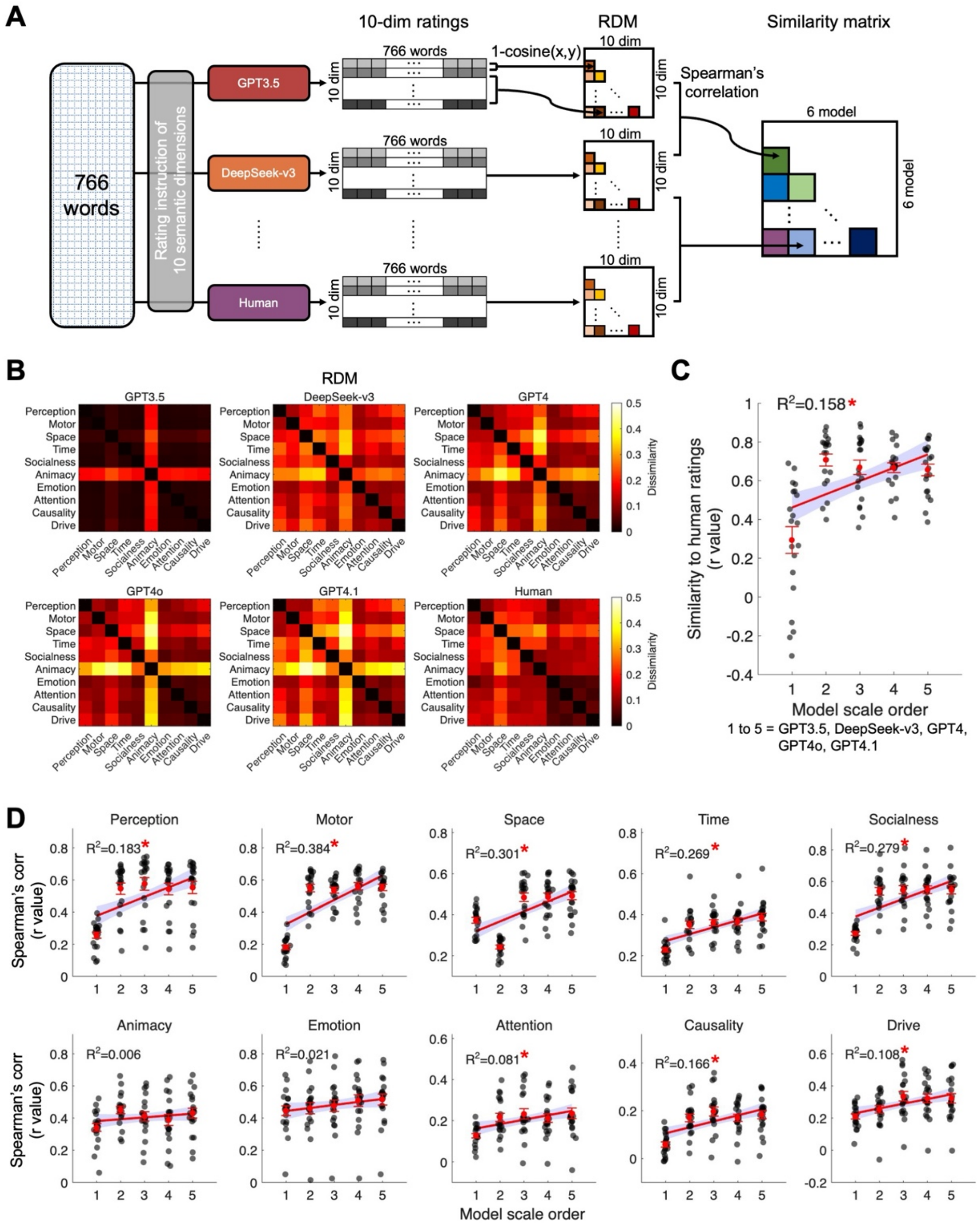

**Figure 2. Similarity between LLMs and humans in semantic rating patterns.** (**A**) Representational similarity analysis (RSA) was conducted on rating patterns between LLMs and humans. 766 words were rated by 5 LLMs and 20 humans on 10 semantic dimensions. Each rating formed a 10-D profile to compute a representational dissimilarity matrix (RDM) showing pairwise dissimilarities based on semantic patterns. RDMs from LLMs and humans were then compared. (**B**) LLM and Human RDMs across 10 semantic dimensions. Dissimilarity between dimensions was calculated as the cosine distance across the entire set of words. (**C**) Linear relationships between the LLM scale (i.e., number of parameters: 1 to 5 represent GPT3.5,

DeepSeek-v3, GPT4, GPT4o and GPT4.1, respectively) and similarities to human RDMs. *, $p < 0.05$ ($F$-test). (**D**) The linear relationships between LLM scale and their similarity to human ratings for each dimension. Scatter grey dots represent LLM's similarity to ratings from individual human raters. Red dots and error bars denote the group mean and standard error of each LLM-human similarity. The solid lines represent the fitted linear regression model, and the shaded areas indicate the 95% confidence interval around the fitted line. Red asterisks indicate significance of linear regression models ($p < 0.05$, $F$-tests).

**Human and LLM semantic spaces explained NS through different representational structures**

We next examined how human- and LLM-derived semantic spaces explained speaker-listener NS, which we used here as a measure of human-shared neural semantics. For each LLM and for the human ratings, we first estimated speaker-induced neural activity separately for each semantic dimension, then used those estimates to predict listeners' neural responses, thereby quantifying semantic NS. When all ten dimensions were modeled jointly, all five LLMs and the human ratings yielded significantly higher overall prediction than the randomized semantic baselines ($t_{(21)}$ range = 10.7-11.3, $p$ range = $1.3 \times 10^{-13}$ to $1.5 \times 10^{-13}$) (Fig. 3A). Because the randomized-rating baseline preserved the timing of content-word occurrences while only disrupting their semantic rating values, the above-chance NS prediction cannot be attributed simply to the temporal distribution of words in the narrative. Further, this overall NS strength did not differ significantly across models and human ($F_{(5,126)} = 1.26$, $p = 0.28$) (Fig. 3A). Thus, at the aggregate level, both human- and LLM-derived semantic spaces captured significant speaker-listener NS beyond simple onset time of the words and their acoustic and phonological features.

At the spatiotemporal level, the resulting NS profiles were broadly similar across humans and LLMs (Fig. 3B). Spatially, strong NS was concentrated between the speaker's right IFG and parietal areas, STG/HG, and the listeners' bilateral STG/HG. Temporally, NS extended from approximately -1200 to 400 ms, peaking at -250 ms (Fig. 3C, left panel), indicating speaker-leading NS. Peak latencies did not differ significantly across models ($F_{(5,126)} = 0.08$, $p = 0.995$) (Fig. S3, left panel). A listener-listener control NS analysis showed significant synchronization from -650 to 500 ms with a peak at -10 ms (Fig. 3C, right panel), again with no effect of model ($F_{(5,126)} =$ 0.02, $p = 0.999$) (Fig. S3, right panel), confirming that the pronounced speaker-leading lag was specific to speaker-listener NS rather than a generic property of the acoustic analysis or due to the speaker hearing their own voice.

Similar overall NS strength between LLMs and humans, however, did not imply they have a similar pattern of neural semantic organization. Pairwise multivariate classification based on spatial NS patterns distinguished most model pairs above chance (permutation $p$ values ranging from $7.5 \times 10^{-4}$ to 0.0497), with the exceptions of GPT3.5 versus human and GPT4.1 versus human (both $p = 0.0836$) (Fig. 3D, left panel). Classification based on temporal NS patterns was weaker, reaching significance only for GPT3.5 versus DeepSeek-v3, GPT4 and GPT4o, for GPT4o versus GPT4.1, and for GPT4.1 versus human (permutation $p$ values ranging from 0.0154 to 0.0369) (Fig. 3D, right panel). When combining both spatial and temporal NS patterns, humans showed significantly distinct NS patterns compared to all LLMs (all permutation $p$ values = $1.0 \times 10^{-4}$). These results indicate that, despite similar overall explanatory power, there are systematic differences in the multivariate structure of human- and LLM-derived NS patterns.

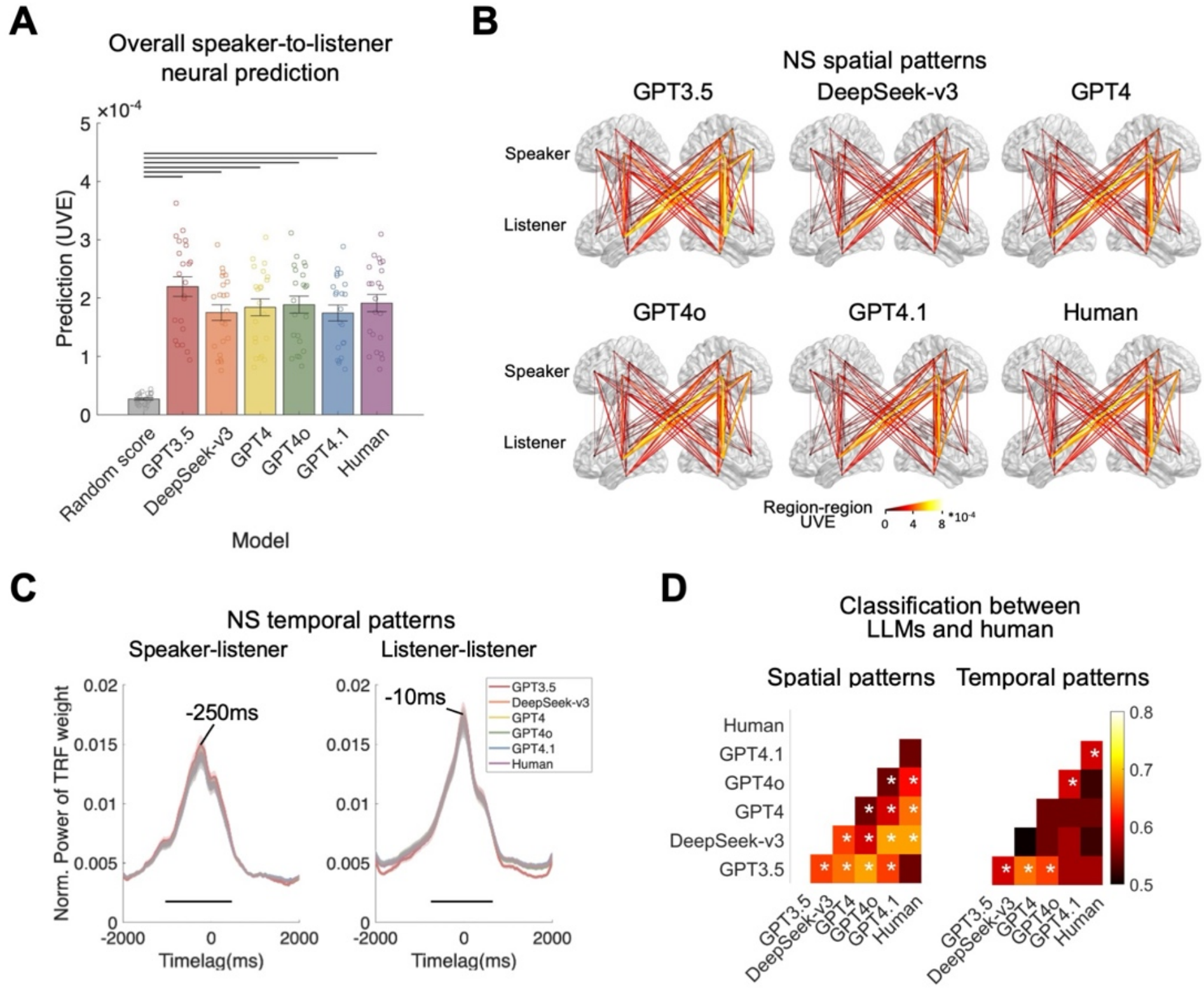

**Figure 3**. **Semantic neural synchronization (NS) patterns derived from LLMs and humans.** (**A**) Unique variance explained (UVE) by LLMs and humans in explaining speaker-listener NS. Random scores serve as the control baseline. Bars show mean ± SEM; circles indicate individual speaker-listener pairs; horizontal lines mark significant pairwise differences. (**B**) Spatial patterns of semantic NS across speaker-listener region pairs. Line color and thickness indicate the prediction strength (i.e., UVE) of region pairs that showed significantly higher prediction than the random score (FDR-corrected $p < 0.05$). (**C**) Normalized TRF-weight power for speaker-listener and listener-listener NS, averaged across dimensions and region pairs. Horizontal lines mark time windows that exceed phase-randomized baselines. (**D**) Classification accuracy for distinguishing models based on spatial and temporal NS patterns. Asterisks indicate above-chance classification accuracy based on permutation tests (FDR-corrected $p < 0.05$).

We thus examined dimension-based spatial NS patterns more directly to reveal human-LLM convergence and divergence using RSA (Fig. 4A). This approach takes into account all ten semantic dimensions in the representational modeling instead of aggregating them or examining them individually. We found that the semantic representational NS similarity differed significantly across models ($F_{(5,126)} = 5.89$, $p = 6.4 \times 10^{-5}$). Human-LLM similarity also differed across LLMs ($F_{(4,105)} = 8.38$, $p = 6.6 \times 10^{-6}$), with GPT4, GPT4o, and GPT4.1 all showing significantly higher similarity to the human than GPT3.5 (post-hoc $p$ values ranging from $2.7 \times 10^{-5}$ to 0.027) (Fig. 4B, left panel). Moreover, human-LLM similarity increased significantly with model scale ($R^2 = 0.214$, $F_{(1,108)} = 30.7$, $p = 2.2 \times 10^{-7}$) (Fig. 4B, right panel). Embedding the dimensional spatial NS patterns into a common space using multidimensional scaling and Procrustes alignment showed that Space and Animacy occupied relatively distinct positions (Fig. 4C). Quantifying distance to the human in this aligned space showed a significant effect of LLM ($F_{(4,105)} = 10.1$, $p = 6.2 \times 10^{-7}$; Fig. 4D), where GPT3.5 was farther from the human than GPT4, GPT4o and GPT4.1 (post-hoc $p$ values ranging from $9.9 \times 10^{-8}$ to $2.9 \times 10^{-3}$), and DeepSeek-v3 was farther than GPT4.1 ($p = 6.1 \times 10^{-3}$). Across dimensions, distances also differed strongly ($F_{(9,210)} = 51.4$, $p = 2.5 \times 10^{-48}$), with Animacy showing the greatest divergence from the human across LLMs (Fig. 4E); there were also significant differences between LLMs for Perception, Time, Animacy and Emotion ($F_{(4,105)}$ values = 3.4-16.6, $p$ values = $1.4 \times 10^{-9}$ to 0.029). Thus, although both humans and LLMs captured the semantic structure that explained NS, their dimension-based neural representational geometries diverged in systematic and model-scale-dependent ways.

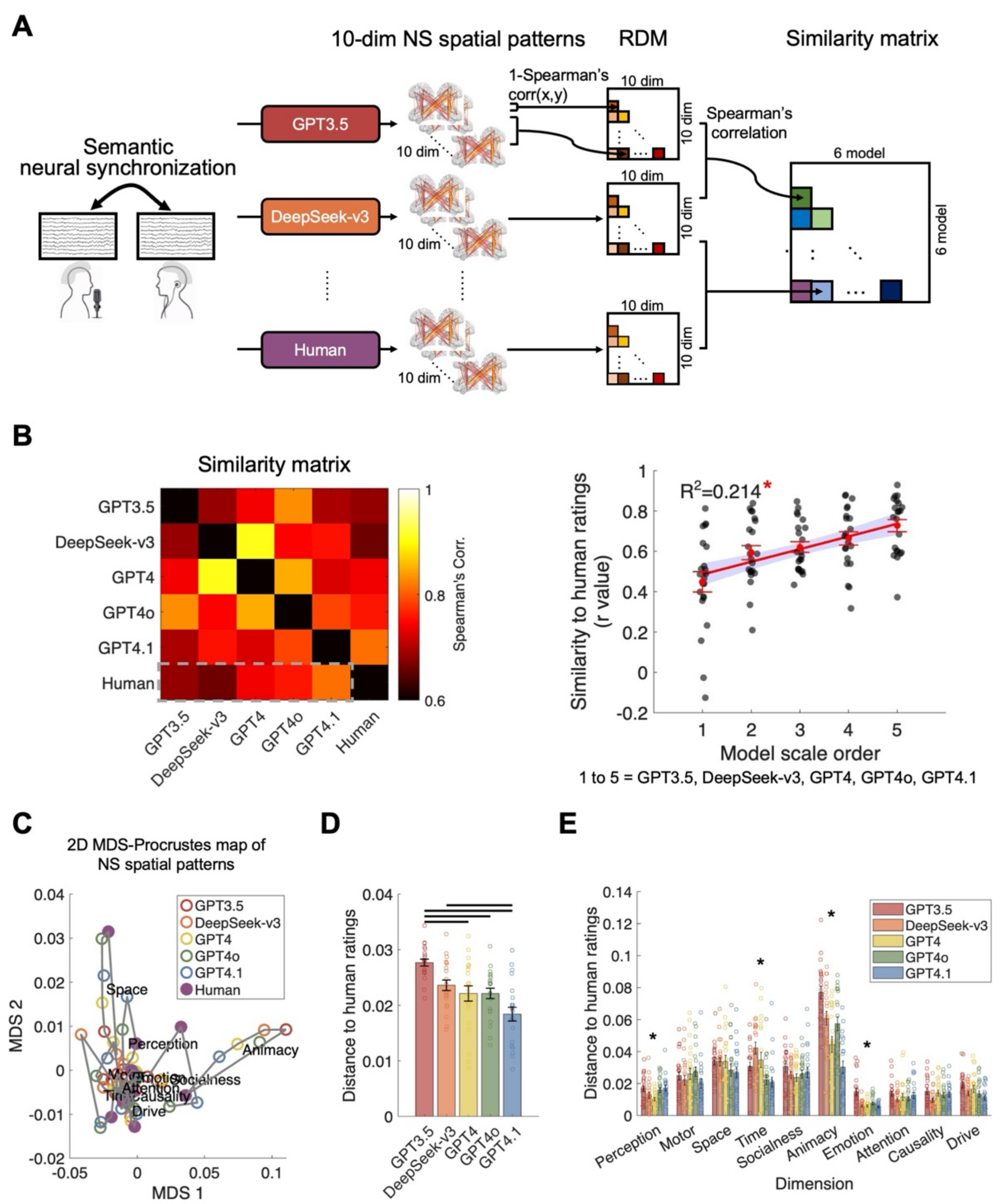

**Figure 4**. **Human-LLM similarity in dimension-based semantic NS patterns. (A)** Representational similarity analysis (RSA) of dimensional NS spatial patterns. For each model, NS patterns were organized as a region-pair × semantic-dimension matrix, and an RDM was computed across the ten semantic dimensions. Pairwise correlations between model RDMs yielded a model-level similarity matrix. (**B**) **Left**, RDM similarity among humans and LLMs with human-LLM similarities highlighted by a dashed box. **Right**, human-LLM NS representation similarity increased with model scale. Scatter grey dots represent human-LLM similarity for each speaker-listener pair. Red dots and error bars denote the group mean and standard error of each LLM. The solid lines represent the fitted linear regression model, and the shaded areas indicate the 95% confidence interval around the fitted line. *, $p < 0.05$. (**C**) MDS-Procrustes embedding of dimensional NS patterns, showing the relative positions of the ten semantic dimensions for both humans and LLMs in a common representational space. (**D**) Overall distance between each LLM and human ratings in the aligned NS space, averaged across dimensions. Bars show mean ± SEM; horizontal lines indicate significant pairwise differences. (**E**) Dimension-specific distance between each LLM and human. Asterisks indicate significant differences among LLMs for a given dimension ($p < 0.05$, one-way ANOVA).

**Dimension-dependent human-LLM semantic NS overlap**

We next quantified the extent to which LLM-derived semantic NS overlapped with the human-derived semantic NS pattern and identified which dimensions captured LLM-specific variance beyond the human semantic model. For each LLM-derived semantic dimension, we estimated its additional contribution to NS after controlling for the full 10-dimensional human semantic model. This additional contribution indexes the non-overlapping variance explained by the LLM dimension that is not captured by human-rated semantic dimensions (i.e., degree of human-LLM divergence).

The extra contribution differed markedly across models and dimensions (Fig. 5A&B, left panel). GPT3.5 showed significant extra contribution for all dimensions ($t_{(21)} = 7.5\text{-}12.3$, $p = 5.5 \times 10^{-10}$ to $3.9 \times 10^{-7}$) except Animacy ($t_{(21)} = -12.7$, $p = 0.999$). DeepSeek-v3 showed significant extra contribution only for Emotion ($t_{(21)} = 11.4$, $p = 5.5 \times 10^{-10}$). GPT4 showed significant extra contribution for Emotion ($t_{(21)} = 11.5$, $p = 5.5 \times 10^{-10}$) and Attention ($t_{(21)} = 4.7$, $p = 1.7 \times 10^{-4}$). GPT4o showed significant additional contributions to Socialness, Emotion, Causality, Drive and Attention ($t_{(21)} = 4.8\text{-}11.9$, $p = 5.5 \times 10^{-10}$ to $1.3 \times 10^{-4}$). GPT4.1 showed significant extra contribution for Emotion ($t_{(21)} = 11.7$, $p = 5.5 \times 10^{-10}$) and Causality ($t_{(21)} = 4.7$, $p = 1.6 \times 10^{-4}$). Thus, LLM-derived semantic dimensions diverged from the human semantic model in a model-specific, non-overlapping way and varied sharply across dimensions, especially in social-affective ones.

We then quantified the degree of convergence between each LLM and the human semantic model (Fig. 5B, right panel & 5C). Convergence differed significantly across LLMs ($F_{(4,105)} = 121.8$, $p = 1.6 \times 10^{-38}$), with GPT4.1 and DeepSeek-v3 showing the highest convergence, GPT4o and GPT4 intermediate, and GPT3.5 the lowest (Fig.

5B, right panel). Convergence also increased significantly with model scale ($R^2 = 0.390$, $F_{(1,108)} = 70.6$, $p = 1.9 \times 10^{-13}$; Fig. 5B, right panel). At the level of individual dimensions, LLMs differed significantly for every dimension except Emotion (all other dimensions: $F_{(4,105)} = 46.3$-$341.3$, $p = 3.3 \times 10^{-58}$ to $2.4 \times 10^{-22}$; Emotion: $F_{(4,105)} = 0.24$, $p = 0.91$; Fig. 5C). Emotion also showed the least degree of convergence with humans across LLMs. Increased model scale predicted greater convergence for all dimensions ($F_{(1,108)}$ ranged from 13.5 to 177.2, $p$-values range from $1.6 \times 10^{-23}$ to $4.6 \times 10^{-4}$) except Emotion ($F_{(1,108)} = 0.17$, $p = 0.68$) and Causality ($F_{(1,108)} = 0.59$, $p = 0.49$) (Fig. 5C). Together, these analyses show that human-LLM overlap in semantic NS was partial and that both the shared fraction and the model-specific remainder varied strongly with semantic dimension.

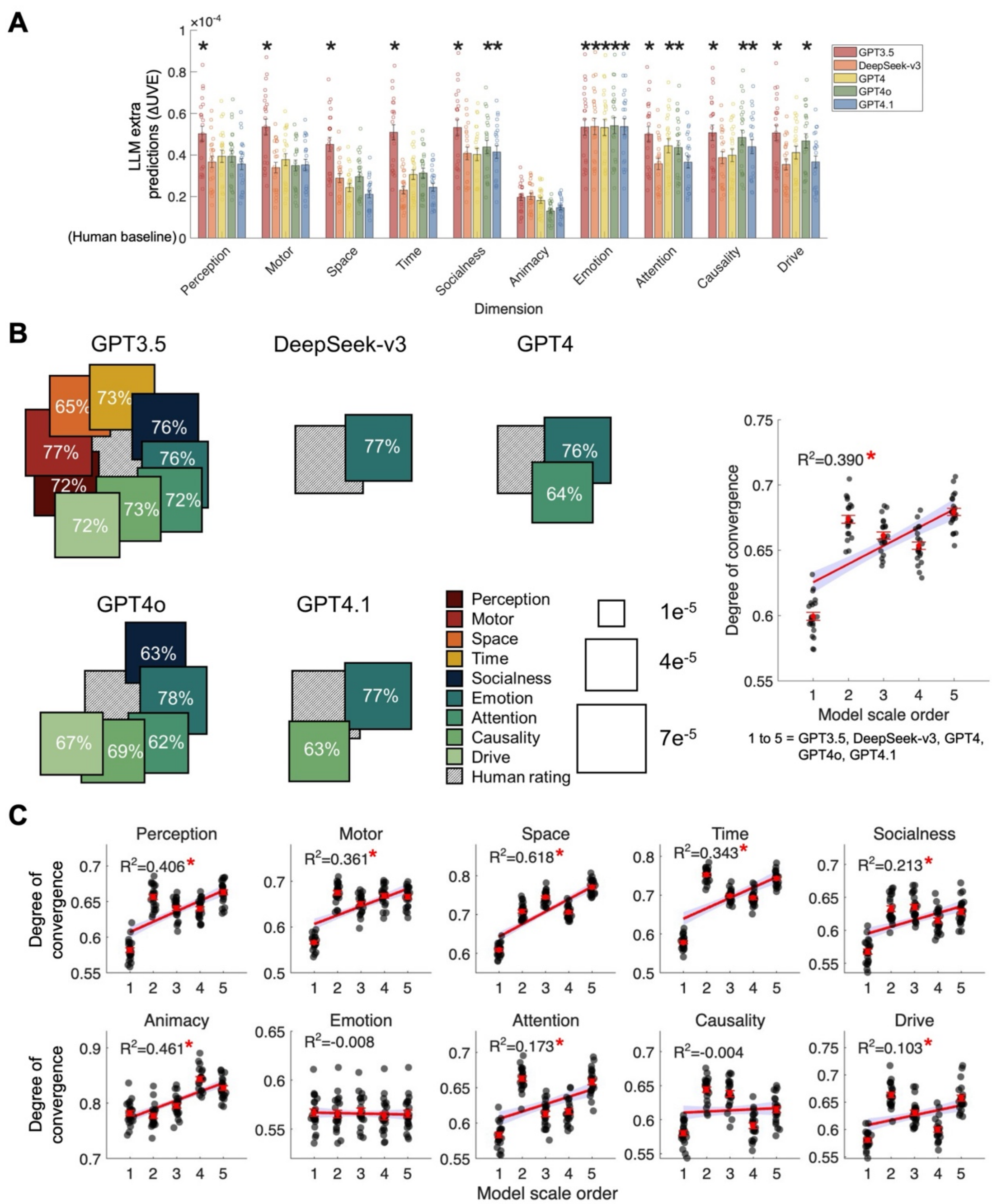

**Figure 5. Shared and distinct components of human- and LLM-derived semantic NS across dimensions.** (**A**) Extra variance in NS explained by each LLM-derived semantic dimension beyond the full human semantic model. Values above zero indicate LLM-specific variance not captured by human ratings; asterisks indicate significant extra contributions. (**B**) **Left**, schematic summary of LLM convergence and divergence NS components. The central grey squares represent the variance explained by the human semantic model. The surrounding colored squares represent significant additional LLM-specific contributions. Square size indicates NS magnitude; percentages indicate extra contribution relative to the human model. **Right**, degree of human-LLM convergence in semantic NS, defined as $R^2(\text{human})/R^2(\text{model} + \text{human})$. Values closer to 1 indicate greater convergence with the human semantic model. Mean convergence across dimensions increased with model scale. Scatter grey dots represent human-LLM similarity for each speaker-listener pair. Red dots and error bars denote the group mean and standard error of each LLM. The solid lines represent the fitted linear regression model, and the shaded areas indicate the 95% confidence interval around the fitted line. The red asterisk represents the significance of the linear regression model. (**C**) Relationship between model scale and convergence for each semantic dimension. Model scale is indexed from 1 to 5, corresponding to GPT-3.5, DeepSeek-v3, GPT-4, GPT-4o, and GPT-4.1. Scatter grey dots indicate individual speaker-listener pairs; red lines and shaded areas show fitted linear trends with 95% confidence intervals; asterisks indicate significant model-scale effects.

**Human-LLM shared semantic geometry was temporally and spatially structured**

The preceding analyses showed that LLM-derived semantic NS partially overlapped with human-derived semantic NS, and that this overlap increased with model scale. We further examined where and when this human-LLM shared and distinct semantic geometry was expressed in space and time. To do so, we applied a searchlight RSA to dimension-resolved NS patterns across time lags and speaker-listener region pairs (Fig. 6A). For each model, we constructed RDMs across the ten semantic dimensions using the corresponding temporal or spatial NS patterns, and then quantified human-LLM similarity by correlating each LLM-derived RDM with the human-derived RDM.

In the temporal domain, larger LLMs showed stronger similarity to human-derived semantic representations across time windows (Fig. 6B). Specifically, the GPT family (i.e., GPT4, GPT4o and GPT4.1) showed a strongly similar temporal profile, with human-LLM similarity peaking at approximately -730 ms. DeepSeek-v3 showed modest similarity and a similar peak time lag. By contrast, GPT3.5 showed substantially lower similarity and an earlier peak at approximately -920 ms. Thus, model-scale convergence was not limited to overall NS magnitude or rating-level similarity; it was also evident in the timing of the shared dimensional semantic geometry between human and LLM-derived NS patterns.

In the spatial domain, human-LLM similarity also increased with model scale and was distributed across multiple speaker-listener region pairs (Fig. 6C). The strongest model-scale effects were observed in region pairs involving the speaker's bilateral frontotemporal-parietal regions, along with listeners' widespread left-hemispheric regions and right parietal-frontal areas. These findings indicate that human-LLM shared semantic representations were expressed through distributed speaker-listener neural pathways with broad timing windows that peak at about -700ms. Together, the temporal and spatial RSA results show that larger LLMs more closely approximated the spatiotemporal geometry of human shared semantic NS.

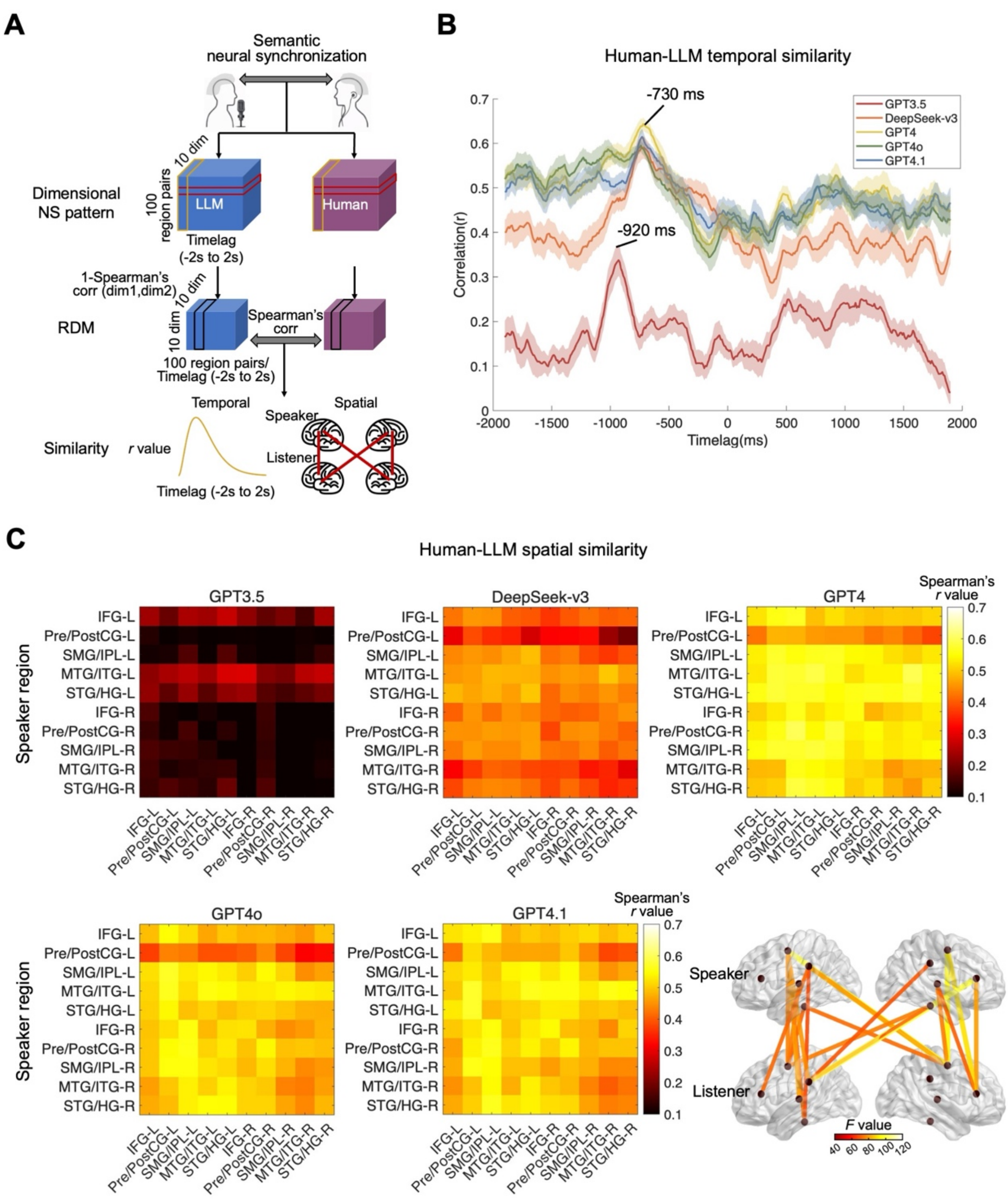

**Figure 6. Spatiotemporal representational similarity between human- and LLM-derived semantic NS.** (**A**) Searchlight RSA framework for quantifying human-LLM similarity in dimensional NS patterns. For each model, TRF weights for 10 semantic dimensions were used to construct RDMs, separately for each time point and each speaker-listener region pair. Human-LLM similarity was computed as the Spearman's correlation between corresponding human and LLM RDMs. (**B**) Temporal profile of human-LLM similarity. Lines show mean ± SEM similarity between human ratings and each LLM across time lags. Larger LLMs showed higher similarity, peaking around -730 ms; GPT3.5 showed distinctly lower similarity with an earlier peak around -920

ms. (**C**) Spatial distribution of human-LLM similarity across speaker-listener region pairs. Matrix plots show the similarity between LLMs and humans. The lower-right brain map highlights the 20 region pairs with the strongest model-scale effects, as indicated by linear regression $F$ values.

**Semantic NS predicted individual differences in narrative comprehension**

Finally, we tested whether dimension-resolved semantic NS was behaviorally relevant by assessing its ability to predict individual differences in listeners' comprehension of the speaker's stories. When all ten semantic dimensions were combined, NS patterns derived from each of the five LLMs and from human ratings significantly predicted comprehension performance ($R^2$ range = 0.340 to 0.378; permutation-based $p = 0.008$; Fig. 7A), but their overall comprehension prediction did not differ ($p$s > 0.05). Thus, semantic NS captured by both human- and LLM-derived ratings was not only neurally significant but also informative about listeners' understanding.

Dimension-specific analyses showed widespread but model-dependent predictive effects (Fig. 7B). Across both LLMs and human ratings, NS from all ten semantic dimensions significantly predicted comprehension above chance (permutation-based $p$-values ranging from 0.0040 to 0.0344). For GPT3.5, DeepSeek-v3, and GPT4o, Animacy showed stronger predictive performance than all other dimensions (permutation-based $p$-values ranging from 0.0018 to 0.0029). For GPT4.1, Perception showed the strongest predictive performance ($p = 0.0015$). For human ratings, Animacy predicted comprehension better than most dimensions, except Perception, Socialness, and Attention (significant differences: $p$-values ranging from 0.0066 to 0.0141; non-significant differences: $p$-values ranging from 0.0584 to 0.1950). Together, these findings show that semantic NS robustly predicted listeners' comprehension, but the dimensions most informative for prediction differed across humans and LLMs.

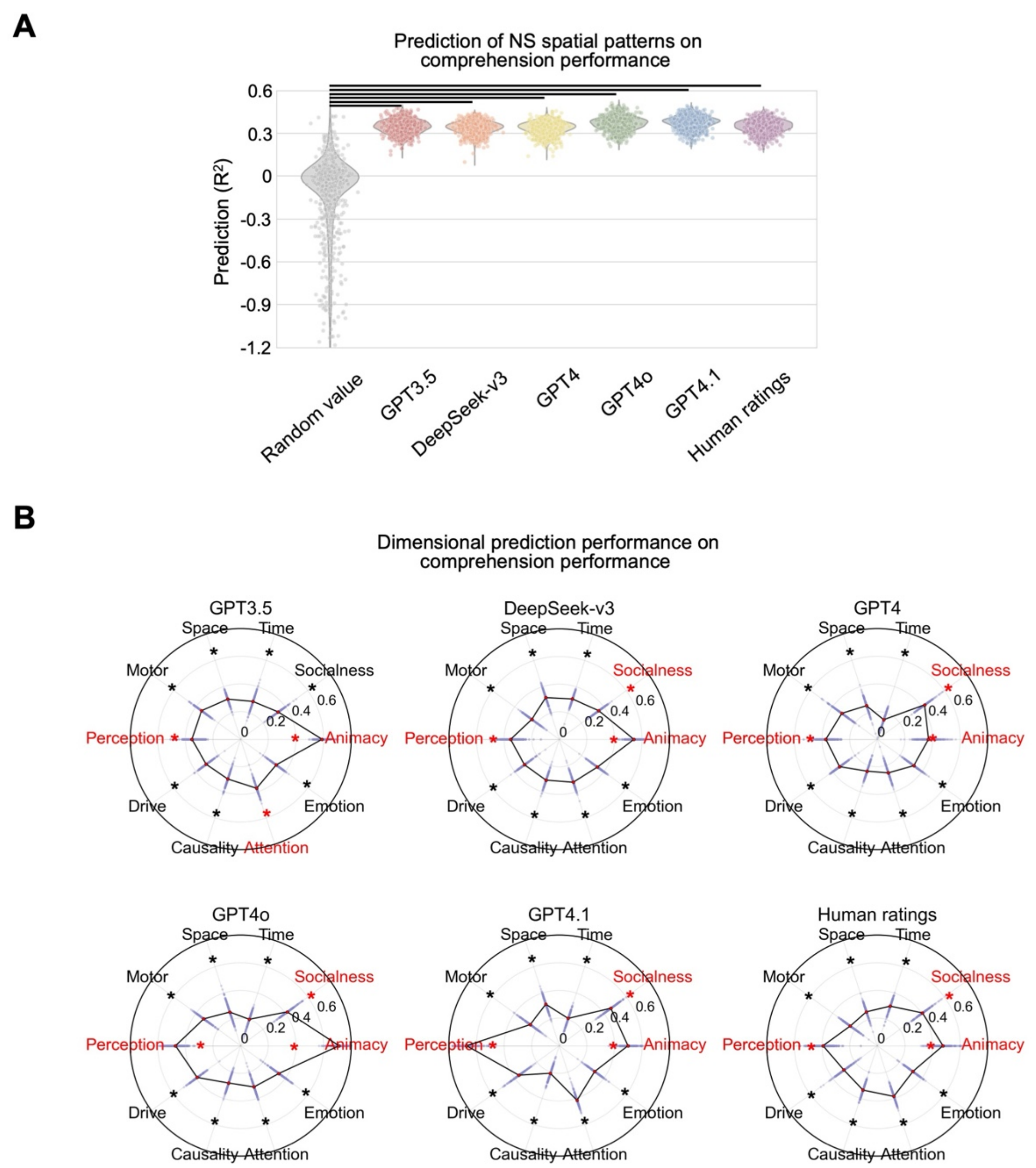

**Figure 7**. **Semantic NS patterns predicted listeners' story comprehension.** (**A**) Prediction of listeners' comprehension from overall semantic NS patterns combining all 10 dimensions. Dots indicate $R^2$ values from 10,000 bootstrapped 10-fold cross-validation runs; horizontal lines mark significant pairwise differences. (**B**) Dimension-specific prediction performance for LLM- and human-derived semantic NS. Asterisks indicate predictions that are significantly above the permutation baseline (after correction for multiple comparisons). The three best-performing dimensions were highlighted in red labels and asterisks.

## Discussion

Our findings provide a dimension-resolved account of shared semantics in naturalistic narrative speech and use this framework to assess human-LLM convergence and divergence in explaining interpersonal semantic representations. Combining storytelling-listening pseudo-hyperscanning MEG with interbrain neural synchronization (NS) modeling, we examined whether semantic alignment between a speaker and listeners can be characterized as a multidimensional structure. Ten semantic dimensions derived from both human and LLM ratings explained speaker-listener NS beyond word onset and across acoustic and phonological controls, indicating that shared neural semantics cannot be reduced to general word representations and lower-level speech features. Although LLM-derived ratings achieved overall NS prediction comparable to human ratings, the semantic dimensions supporting this alignment showed systematic differences in representational geometry. Larger models showed stronger convergence with human semantic ratings and human-derived NS patterns, but this convergence was partial and varied across dimensions, with strong divergences observed for animacy, emotion and socialness. Finally, semantic NS predicted individual differences in listeners' story comprehension, linking dimension-resolved NS to behavioral understanding.

**Comparable NS strengths do not indicate an equivalent shared semantic representation structure**

LLM-derived semantic ratings explained speaker-listener NS at a level comparable to human ratings overall, yet differed from humans in the organization of the semantic dimensions that drove this alignment. This dissociation is theoretically insightful because it shows that strong neural prediction alone is insufficient evidence that an artificial model represents shared semantics in a human-like way. A model may capture broad semantic regularities that are useful for predicting human-shared neural alignment while differing from humans in how specific semantic dimensions are organized, weighted, and mapped onto speaker-listener neural pathways.

This distinction refines how human-LLM alignment should be evaluated in cognitive neuroscience. Prior work shows that LLM representations can predict neural responses during naturalistic language processing and capture aspects of speaker-listener shared linguistic content (Antonello et al., 2023; Bonnasse-Gahot & Pallier, 2024; Zada et al., 2026). Our findings suggest that overall prediction accuracy provides only a coarse measure of human-model alignment. A stricter test is whether model-derived semantics reproduce the dimensional geometry of human-shared semantic NS, including the organization of semantic dimensions and the spatiotemporal NS patterns that support them. By this criterion, LLMs showed significant but incomplete convergence with humans.

Thus, the findings support a selective account of human-LLM alignment and, more broadly, of human-shared semantic representations. Although LLMs captured substantial components of the semantics that support speaker-listener NS, their convergence with humans was incomplete. This pattern suggests that shared neural semantics includes both distributionally learnable structure, as represented by LLMs, and dimensions more closely tied to human experience.

**Animacy, emotion, and socialness reveal the strongest human-LLM divergence**

The strong human-LLM divergences emerged in animacy, emotion, and socialness. These dimensions are theoretically diagnostic because they are not merely lexical or distributional properties. They are closely linked to agency, affective experience, interpersonal relations, and social inference. This divergence supports the view that human semantic representations are shaped not only by linguistic co-occurrence or text regularities but also by embodied and social experience (Gallese & Lakoff, 2005; Martin, 2007; Pulvermüller & Fadiga, 2010; Vigliocco et al., 2011). By contrast, most LLMs derive semantic structure primarily from statistical regularities in text, which may allow them to approximate many semantic associations while leaving experientially grounded dimensions only partially captured (Du et al., 2025; Xu et al., 2025).

Emotion showed particularly weak convergence with human-derived semantic NS and showed little improvement with model scale. This suggests that LLMs may approximate emotion-related meanings at the distributional level but differ from humans in how affective meaning contributes to shared neural semantics. Human emotion semantics are shaped by bodily states, motivational relevance, interpersonal context, and intentional meaning. These experiential factors are central to the development of affective conceptual representation (Kousta et al., 2011; Vigliocco et al., 2014), but they are not directly available to text-trained language models. This may explain why emotion remained difficult for LLMs to align with human shared neural semantics.

Animacy also diverged strongly from human-derived NS patterns. In communicative contexts, animacy often conveys information about agency, intention, mental states, and social roles. Human listeners may use animacy cues to infer who is acting, who has goals, whose perspective matters, and how events should be interpreted. This interpretation aligns with evidence that animate entities are represented in partially specialized semantic networks (Fairhall & Caramazza, 2013; Hauptman & Bedny, 2025) and with broader work on mental-state attribution and social cognition (Frith & Frith, 2010; Saxe & Kanwisher, 2003). LLMs can learn many animacy-related properties and regularities from text, but these representations may not fully reproduce the socially grounded inferences that shape human semantic alignment.

Socialness provides a particularly informative test case for human-LLM convergence because social meaning is difficult to reduce to statistical regularities. Human social semantics depends on conceptual knowledge about people, interpersonal relations, norms, intentions, and context-appropriate behavior, and naturalistic neuroimaging evidence suggests that social and semantic cognition are supported by partially overlapping but distinct neural systems (Binney & Ramsey, 2020; Thye et al., 2024). Although LLMs can learn many text-based regularities in social language, their social-language understanding remains uneven across domains, and human-like performance on some theory-of-mind tasks does not necessarily imply human-like social representation (Choi et al., 2023; Strachan et al., 2024). The observed divergence in socialness may therefore indicate that LLMs capture distributional associations in social content without fully aligning with the experience-grounded and interaction-sensitive structure of human social semantics. Together with divergences in animacy and emotion, this pattern suggests that the strongest human-LLM gaps arise where meaning depends most strongly on agency, affect, and interpersonal experience (Vigliocco et al., 2014; Xu et al., 2025).

**Model scale selectively improves convergence**

A second important finding is that human-LLM convergence generally increased with model scale or capability. Larger LLMs showed greater similarity to human semantic ratings, stronger similarity to human-derived dimensional NS patterns, and greater overlap with the structure of the human semantic representation. This pattern aligns with evidence that larger language models show stronger alignment with human behavioral and neural responses during naturalistic language processing (Antonello et al., 2023; Bonnasse-Gahot & Pallier, 2024; Gao et al., 2025).

However, model scaling did not yield uniform convergence across semantic dimensions. At the behavioral rating level, larger models showed stronger similarity to human ratings for most dimensions, but not for Emotion or Animacy. At the neural level, scaling increased human-LLM convergence for most dimensional NS patterns, but this effect was absent for Emotion and Causality, and Animacy remained the most divergent dimension in the shared NS geometry. This uneven profile suggests that scaling improves alignment for dimensions that are more recoverable from linguistic distributional structure, including aspects of temporal and perceptual meaning, consistent with evidence that larger language models better predict human neural responses and can encode structured spatial and temporal information (Antonello et al., 2023; Gurnee & Tegmark, 2024). In contrast, scaling appears less sufficient for dimensions that depend more strongly on affective grounding, agency-related interpretation, social inference, or event-level causal structure. This interpretation is consistent with evidence that text-only LLMs show weaker alignment with human concepts when sensory-motor grounding is required, that emotion contributes to semantic representation, and that LLMs show uneven performance in social-pragmatic and theory-of-mind tasks (Choi et al., 2023; Strachan et al., 2024; Vigliocco et al., 2014; Xu et al., 2025). Larger models are therefore more human-like in some aspects of shared neural semantics, but scaling alone does not eliminate the gap between text-derived semantic structure and human semantic experience.

**Multidimensional shared neural semantics are behaviorally relevant**

The behavioral prediction results further suggest that dimension-resolved semantic NS is functionally meaningful. NS patterns derived from both human and LLM ratings significantly predicted individual differences in listeners' story comprehension, consistent with prior evidence that stronger speaker-listener coupling predicts better communicative understanding and that neural coupling is linked to semantic features of speech (Li et al., 2024; Stephens et al., 2010). These findings argue against the interpretation that semantic NS merely reflects shared auditory input or model-fitting artifacts (Liu et al., 2020). Instead, they suggest that the degree to which semantic dimensions organize speaker-listener neural alignment contains information about how successfully listeners reconstruct narrative meaning.

At the dimension level, predictive effects were widespread but model-dependent. Animacy was especially predictive for several LLMs and for human ratings, whereas Perception was the strongest predictor for GPT4.1. This pattern is consistent with situation-model theories, which propose that narrative comprehension requires tracking agents, goals, causal relations, temporal structure, spatial context, and perceptual details as events unfold (Radvansky & Zacks, 2014; Zwaan & Radvansky, 1998). The strong role of Animacy is also consistent with evidence that animacy is not a simple living-nonliving distinction but involves agency, mobility, and behavioral relevance (Jozwik et al., 2022). Importantly, behavioral relevance and human-LLM divergence are not identical. Animacy was both strongly divergent and strongly predictive, whereas Perception was behaviorally important despite showing less divergence from the human-derived NS geometry. This distinction clarifies that human-LLM divergence identifies where models differ from humans, whereas behavioral prediction identifies which semantic NS patterns explain individual differences in understanding.

**Spatiotemporal semantic NS reflects one-way narrative transmission**

The spatiotemporal profile of semantic NS further constrains the interpretation of shared semantic representations. Across human- and LLM-derived semantic spaces, semantic NS showed a broad temporal window dominated by speaker-leading alignment. The overall semantic NS signal peaked at approximately −250 ms, indicating that speaker-derived semantic activity predicted listeners' later responses. This pattern is consistent with prior evidence that speaker–listener coupling during natural communication unfolds across time and supports successful understanding (Chang et al., 2024; Stephens et al., 2010). A smaller listener-leading component may reflect anticipatory or predictive processes during comprehension, consistent with hierarchical predictive accounts of speech understanding (Caucheteux et al., 2023; Friston, 2005; Stephens et al., 2010).

Spatially, semantic NS was mainly expressed through coupling between the speaker's frontal-temporal-parietal regions and the listeners' temporal cortices. This pattern is consistent with neurocognitive models in which speech production recruits frontal, temporal, and parietal systems for planning, lexical-semantic access, and integration, whereas comprehension relies strongly on temporal-lobe mechanisms for mapping speech input onto meaning (Binder et al., 2009; Friederici, 2011; Hickok, 2012; Hickok & Poeppel, 2007). The present study's contribution lies in demonstrating how semantic dimensions influence the activation of specific speaker-listener pathways.

The searchlight RSA further showed that human-LLM shared semantic geometry was temporally and spatially structured. Importantly, the peak around -730 ms was also speaker-leading, but it indexed a different level of alignment from the -250 ms peak. Whereas the -250 ms peak reflected the strength of overall semantic NS, the -730 ms peak reflected the time at which the dimensional geometry of semantic NS was most similar between human- and LLM-derived models. This longer speaker-leading lag may indicate that the structured organization among semantic dimensions emerges over a broader temporal window than overall semantic coupling strength, possibly reflecting higher-order integration of semantic relations across dimensions. This interpretation is consistent with evidence that LLM-derived linguistic spaces can capture information transmitted from speaker to listener brains during natural communication (Zada et al., 2024). However, because the present study used pseudo-hyperscanning, these dynamics should be interpreted as one-way narrative transmission from speaker to listener rather than as reciprocal dialogue.

**Limitations and future directions**

Several limitations should be noted. First, the pseudo-hyperscanning design modeled one-way narrative transmission rather than real-time interaction. Although this controlled auditory input across listeners, it could not capture turn-taking, adaptation, or mutual prediction. Future hyperscanning studies during natural dialogue should test whether reciprocal interaction affects the same dimension-specific convergence and divergence patterns, especially for socialness and emotion (Hari & Kujala, 2009; Montague et al., 2002; Pickering & Garrod, 2004; Zada et al., 2026). Second, the study used one selected speaker and selected speech-language ROIs. This improved experimental control and reduced model complexity, but limits generalizability across speakers, discourse styles, and broader narrative or social-cognitive networks. Future work should include multiple speakers and extend the analysis to whole-brain networks. Third, LLMs differed not only in estimated scale or capability but also in training data, tuning, alignment procedures, and possible multimodal exposure. Model scale should therefore not be treated as the sole source of human-LLM convergence. Future studies should systematically compare text-only and multimodal models to test whether perceptual or social grounding improves alignment with human semantic NS, especially for perception, emotion, animacy, and socialness (Du et al., 2025; Xu et al., 2025). Finally, the semantic space was limited to ten predefined dimensions and Mandarin content words. Future work should incorporate additional semantic dimensions, contextualized embeddings, discourse-level variables, cross-linguistic comparisons, and individual differences in social cognition, affective traits, language experience, or narrative ability.

## Conclusion

We demonstrate that human-shared semantic representations during narrative speech are multidimensional and measurable through speaker-listener NS. Human- and LLM-derived semantic spaces both explained interbrain NS and predicted comprehension, supporting LLMs as models of shared semantics. However, the overall prediction did not imply identical representational geometry. Larger models showed stronger convergence with human NS, but gaps remained, especially in the dimensions of animacy, emotion, and socialness, which are linked to agency, affect, and interpersonal experience. Linking semantic dimensions, neural alignment, and comprehension provides a framework for evaluating the nature of shared semantics and how well LLMs mirror human semantics in naturalistic narratives.

**Supplementary Materials**

**Table S1.** Working definition of each semantic dimension

| Dimension | Working definition |
|---|---|
| Perception | The extent to which the meaning of a word can easily and quickly trigger corresponding sensory experience (including sight, sound, touch, taste, smell, etc) in your mind. |
| Motor | The extent to which the meaning of a word can easily and quickly trigger corresponding bodily movement or action execution in your mind. |
| Space | The extent to which the meaning of a word relates to spatial information, including location, direction, distance, path, scene, etc. |
| Time | The extent to which the meaning of a word relates to time, including early or late, length, sequence, frequency, etc. |
| Socialness | The extent to which the meaning of a word relates to interpersonal interaction, group dynamics, or social norms. |
| Animacy | The extent to which the meaning of a word relates to humans, living beings, agency, or biological activity. |
| Emotion | The extent to which the meaning of a word relates to positive or negative affective states, feelings, or emotional responses. |
| Attention | The extent to which the meaning of a word relates to attention, focus, salience, or cognitive selection of stimuli. |
| Causality | The extent to which the meaning of a word implies a causal or correlated relationship or agency. |
| Drive | The extent to which the meaning of a word relates to motivation, goals, needs, or desires. |

**Table S2.** Descriptive statistics and representative examples of subjective ratings across the ten semantic dimensions. For each dimension, the table reports the mean, standard deviation (SD), minimum rating (Min), the example word with the minimum rating, maximum rating (Max), and the example word with the maximum rating. English translations of the Chinese example words are provided in parentheses.

| Dimension | Mean | SD | Min | Min example | Max | Max example |
|---|---|---|---|---|---|---|
| Perception | 3.276 | 1.139 | 1.15 | 方式/method | 6.25 | 霓虹灯/neon light |
| Motor | 2.701 | 1.215 | 1 | 年纪/age | 6.45 | 敲门/knock |
| Space | 1.595 | 0.985 | 1 | 头发/hair | 5.95 | 机场/airport |
| Time | 1.423 | 0.700 | 1 | 木头/wood | 6.7 | 小时/hour |
| Socialness | 2.132 | 1.177 | 1 | 衣柜/closet | 6 | 婚姻/marriage |
| Animacy | 3.206 | 1.676 | 1 | 香皂/soup | 7 | 父亲/father |
| Emotion | 0.296 | 0.873 | -3.789 | 哭/cry | 5.15 | 爱/love |
| Attention | 1.536 | 0.483 | 1 | 被子/quilt | 4.8 | 关心/care |
| Causality | 1.423 | 0.438 | 1 | 床/bed | 5.5 | 导致/cause |
| Drive | 1.549 | 0.585 | 1 | 门/door | 5.45 | 希望/hope |

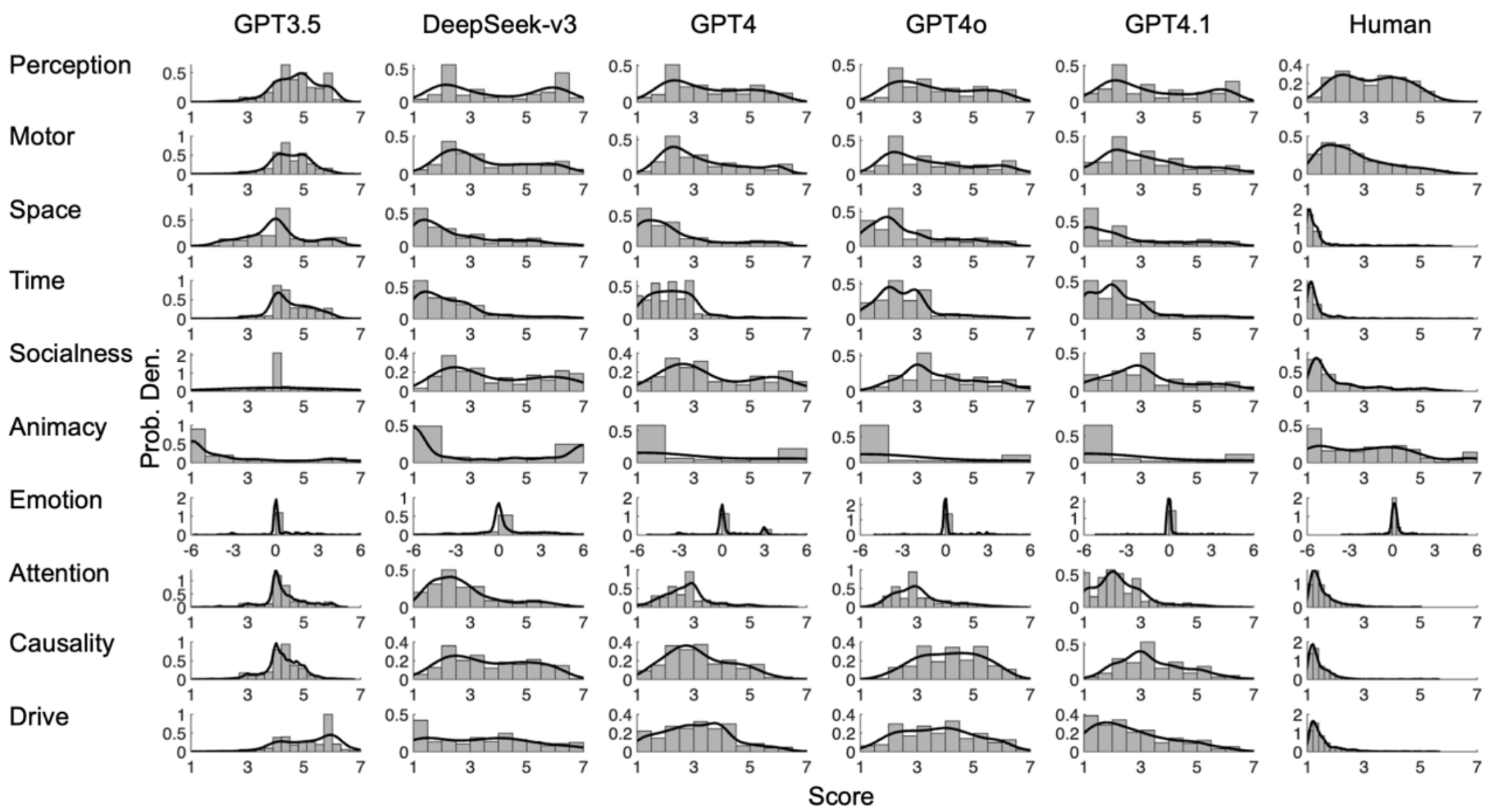

**Figure S1.** Distributions of semantic ratings derived from LLMs and human raters across the ten semantic dimensions. The x-axis represents rating scores, and the y-axis represents probability density. Gray histograms show the empirical rating distributions, and black curves show the fitted density distributions.

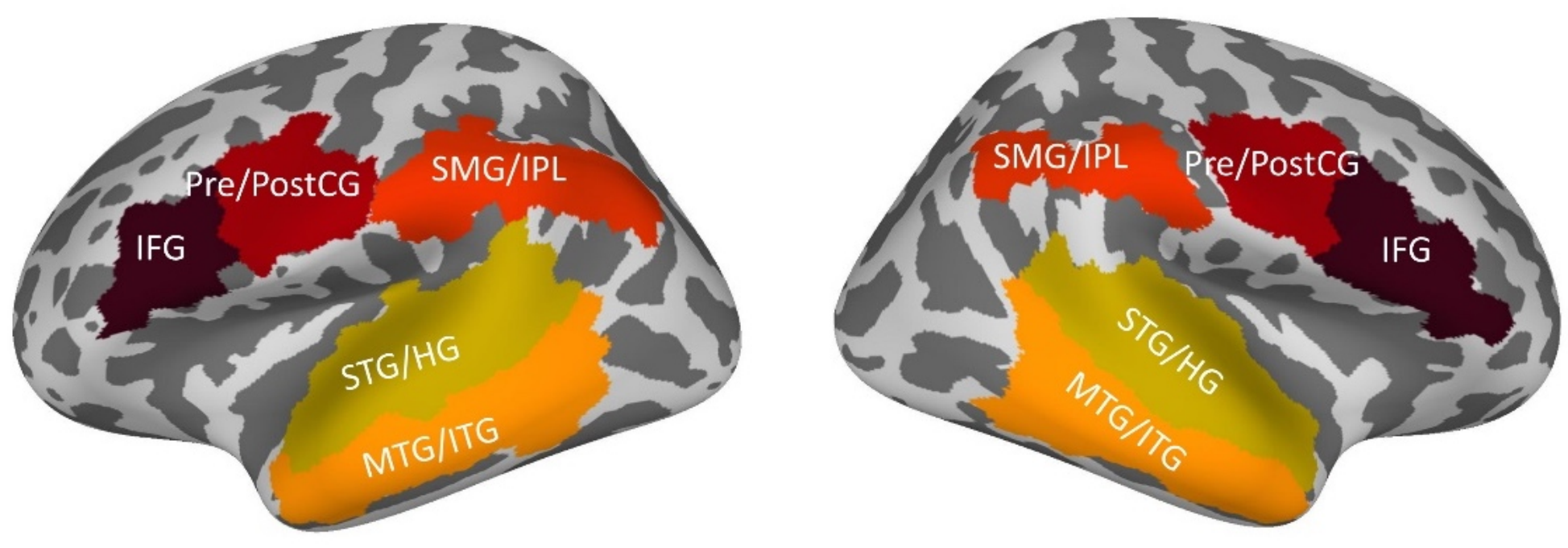

**Figure S2.** Speech and language regions of interest (ROIs) used in the NS analysis. Bilateral cortical ROIs are displayed on brain surface maps and include the inferior frontal gyrus (IFG), precentral/postcentral gyri (Pre/PostCG), supramarginal/inferior parietal cortex (SMG/IPL), superior temporal/Heschl's gyri (STG/HG), and middle/inferior temporal gyri (MTG/ITG). These regions were selected to cover frontal, temporal, sensorimotor, and parietal components of speech production and comprehension networks. This ROI definition is consistent with neurocognitive models of language that emphasize distributed frontal-temporal-parietal systems for speech perception, production, and integration.

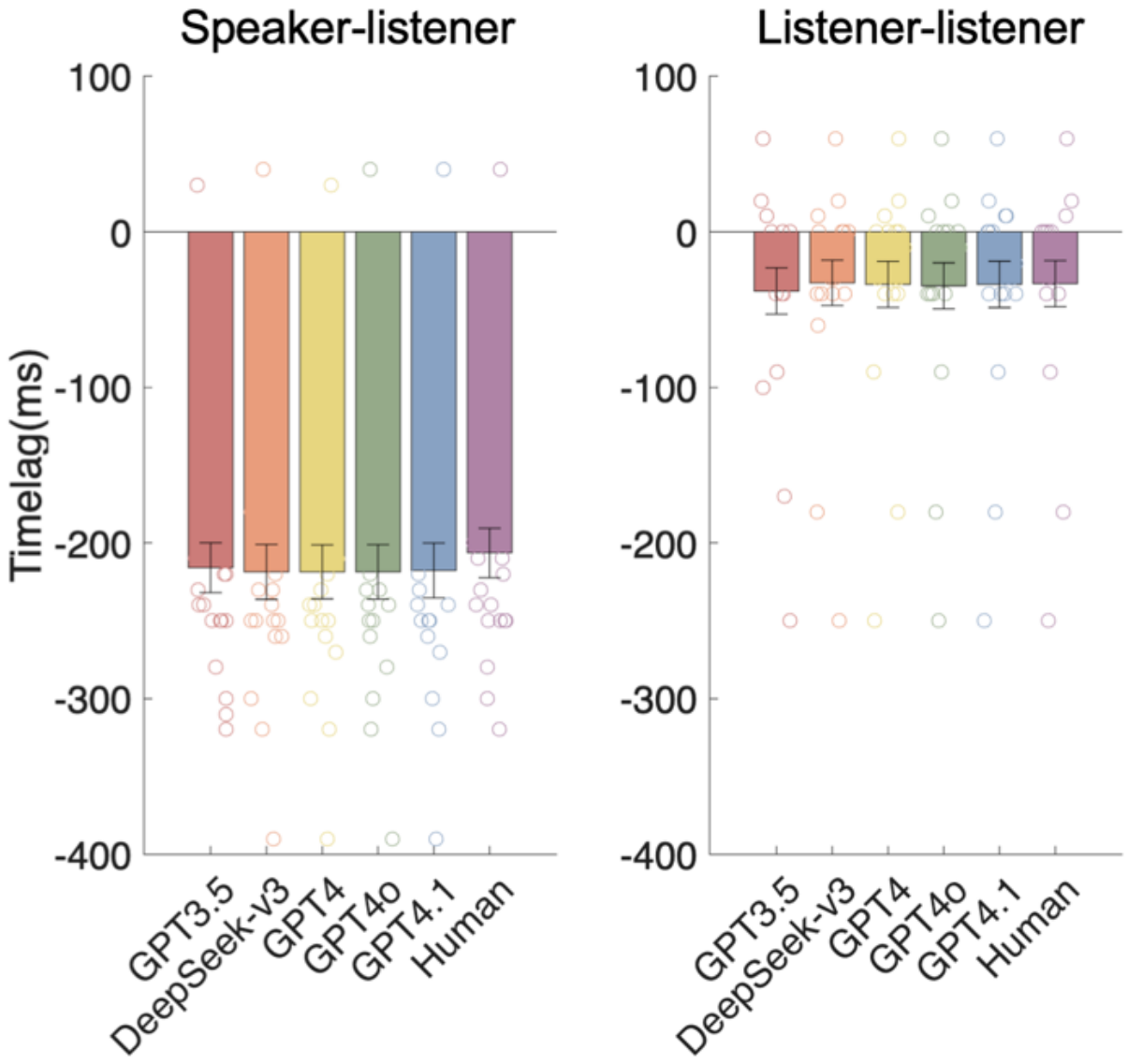

**Figure S3.** Peak latencies of semantic NS for speaker-listener and listener-listener pairs. Each dot represents one pairwise estimate, and bars indicate mean ± SEM. Speaker-listener peak latencies quantify the temporal lag at which speaker-derived activity most strongly predicted listener responses, whereas listener-listener peak latencies provide a control estimate of synchronization across listeners exposed to the same narratives. This comparison distinguishes directional speaker-to-listener alignment from shared stimulus-locked listener responses, consistent with prior work on temporally lagged speaker-listener coupling during natural communication.